\documentclass[a4paper,fleqn,usenatbib,useAMS]{mnras}

\usepackage{comment}
\usepackage{graphicx}
\usepackage{amsmath}	
\usepackage{amssymb}
\usepackage{multicol}       
\usepackage{bm}		
\usepackage{pdflscape}
\usepackage{subfigure}
\usepackage{multirow}

\usepackage{tikz}
\usepackage{amsmath}
\usetikzlibrary{calc}
\usetikzlibrary{decorations.markings}
\usetikzlibrary{shadings, calc, decorations.markings}
\tikzset{->-/.style={decoration={
  markings,
  mark=at position #1 with {\arrow{>}}},postaction={decorate}},
  ->-/.default=0.5,
  }

\usepackage[T1]{fontenc}
\usepackage{ae,aecompl}

\usepackage{newtxtext,newtxmath}

\title[Gravitational radiation by magnetic field]{Gravitational radiation by magnetic field: application to millisecond magnetars}

\author[E. Nazari \& M. Roshan]{
Elham Nazari$^{1}$\thanks{Contact e-mail: \href{mailto: elham.nazari@mail.um.ac.ir}{elham.nazari@mail.um.ac.ir}}
Mahmood Roshan$^{1,2}$\thanks{Corresponding author: \href{mailto: mroshan@um.ac.ir}{mroshan@um.ac.ir}}
\\
$^{1}$Department of Physics, Faculty of Science, Ferdowsi University of Mashhad P.O. Box 1436, Mashhad, Iran \\
$^{2}$School of Astronomy, Institute for Research in Fundamental Sciences (IPM), Tehran, 19395-5531, Iran}

\date{Last updated 2020 June 17; in original form 2020 June 8}

\pubyear{2020}

\begin{document}
\label{firstpage}
\pagerange{\pageref{firstpage}--\pageref{lastpage}}
\maketitle

\begin{abstract}

We investigate the direct contribution of the magnetic field to the gravitational wave generation. To do so, we study the post-Newtonian energy-momentum tensor of the magnetized fluid and the post-Newtonian expansion of the gravitational potential in the wave zone. We show that the magnetic field appears even in the first post-Newtonian order of the multipole moment tensor. Then, we find an explicit relativistic correction containing the magnetic field contribution to the well-known quadrupole formula. As an application of this derivation, we find that the B-field part of the gravitational waves released in the early stages of a millisecond magnetar's life can be as much as one-hundredth of the signals due to the deformed rotating neutron stars. We show that although the event rate of this system is small, the signal would lie in the sensitivity range of the next generation of detectors.

\end{abstract}


\section{Introduction}\label{introduction}

Recently, a new episode in astronomy and astrophysics have been started after detecting the gravitational waves (GWs) generated from the binary black hole and binary neutron star (BNS) mergers by LIGO and Virgo (\citealp{abbott2016observation,abbott2016gw151226,abbott2017gw170104,abbott2017gw170814,abbott2017gw170817}). This new window onto the universe which does not rely on the electromagnetic radiation would bring new information about it.    
There are several astrophysical candidates for sources of GWs, e.g., the gravitational collapse, deformed rotating neutron star (NS), binary systems, quasi-normal modes of a black hole, and stochastic background, for a review on the subject see \cite{sathyaprakash2009physics,bonazzola1994astrophysical}. Among these sources, the coalescence of the binary systems produces the strongest GW, and the current GW detectors can only observe these types of waves. Nevertheless, it is expected by constructing and progressively developing space-based and ground-based interferometers, weaker signals can be discovered in the near future (\citealp{sathyaprakash2012scientific,crowder2005beyond,sato2017status}). 
Furthermore, It is recently shown in \cite{abbott2020gw190412} that the contribution of the higher multipoles is detectable in GW signals by the next generation GW detectors.
Therefore, a more detailed study of GWs emitted from other sources seems to be necessary.

As we know, the deformed NS, as a possible source of GWs, is of interest. This deformation can be induced by several phenomena. For example, the existence of the strong magnetic field inside NS is believed to be flux frozen and independent of the spin of the star (\citealp{bocquet1995rotating,bonazzola1996gravitational,ioka2001magnetic}).
This class of highly magnetized NSs is named magnetar. The Soft Gamma Repeaters (SGRs) and the Anomalous X-ray Pulsars (AXPs) are categorized as magnetar candidates (\citealp{duncan1992formation,thompson1993neutron,woods2006soft}). The order of magnitude of the magnetic field is estimated about $10^{15}\text{G}$ on the surface of these systems (\citealp{melatos1999bumpy,makishima2014possible}).
Moreover, magnetars may also be the remnant of BNS mergers, e.g., see \cite{metzger2018magnetar,margalit2019fast}.
In \cite{usov1992millisecond}, \cite{nakamura1998model}, \cite{kluzniak1998central}, and \cite{wheeler2000asymmetric}, it is shown that even the maximum strength of the magnetic field inside NS can be of the order $\sim10^{18}\text{G}$.
There have been a lot of works on the modeling of the magnetar and the configuration of its magnetic field in the context of the general relativistic magnetohydrodynamics (GRMHD). For instance, in \cite{bonazzola1993axisymmetric}, \cite{bocquet1995rotating}, and \cite{cardall2001effects}, the magnetic field only has the poloidal component, while both toroidal and poloidal magnetic fields are considered in \cite{ioka2004relativistic} and \cite{ciolfi2013twisted}.

In the dipole model for magnetars, the inclination between the magnetic dipole moment and the rotation axis of the magnetar can induce the magnetic ellipticity and consequently produce GW. The evolution of the angle between the magnetic and rotation axes are also investigated (see e.g., \cite{lander2018neutron} and references therein). In fact, the asymmetric magnetic pressure and boundary conditions of the magnetic field at the stellar surface distort the magnetar. Therefore, the rotation of the deformed star around the non-principal axes generates GW, for more detail see e.g., \cite{bonazzola1996gravitational,cutler2002gravitational,tomimura2005new,haskell2008modelling,sieniawska2019continuous}. 
The ellipticities of the relativistic stars induced by a toroidal magnetic field and rotation are also studied in \cite{frieben2012equilibrium}. 
In this particular context, by exploring the Lorentz force on the matter, the shape of the ellipsoid can be found. By obtaining this quantity, the quadrupole moment is then calculated, and finally, GWs in which the effect of the magnetic field is considered can be immediately deduced.
Therefore, the main idea is to obtain the excited GWs by taking into account the influence of the magnetic field on the matter and the creation of an asymmetric structure by it. 
We call it the "indirect" effect of the magnetic field on GWs.

On the other hand,
the super-strong magnetic field around magnetars appears as an asymmetric "mass-energy" distribution. Therefore, in principle, it can "directly" contribute to GWs signals.
Indeed, a strong magnetic field can appear not only in the equation of motion of the system's components and consequently in GWs, as an indirect role, but also in the multipole moment of the source. Therefore, it may directly change the final GW signals.
As we know, GW can be appropriately explained by applying the post-Newtonian (PN) theory to the higher-order corrections (\citealp{poisson2014gravity}).
On other words, in this context, the PN expansion of the multipole moment is constructed from its components, i.e., the energy-momentum tensor, step by step to the required degree of the accuracy. Finally, it forms the PN waveform of the GW signals. 
This formalism is used to study the GW templates to 3.5PN order (see \cite{blanchet2014gravitational} and references therein).
This method is widely applied to study the experimental foundations of general relativity (GR) (\citealp{thorne1971theoreticali,will1971theoreticalii}), gravitational radiations from binary systems (\citealp{blanchet1989post,blanchet1989higher,blanchet1995gravitational}), equation of motion of binary systems (\citealp{hulse1975deep,blandford1976arrival,damour1991orbital}), gravitational radiation reaction (\citealp{chandrasekhar1970212,burke1971gravitational,blanchet2014gravitational}), and so on. To see more applications of this theory, we refer the interested reader to \cite{will2014confrontation}. 
In this context, the famous Einstein quadrupole formula is the leading term of the exact GW signals. As mentioned before, the important step to derive GW is to calculate the PN multipole moment of the source. To do so, one should have enough information about the energy-momentum tensor of the system to the required PN corrections. Therefore, by considering the role of the magnetic field in the PN expansion of the energy-momentum tensor, one can find the direct influence of the magnetic field on GW.

In the present work, by utilizing the results of \cite{nazari2018post} where the PN energy-momentum tensor is obtained in the presence of the electromagnetic field, we derive the B-field contribution to the PN expansion of the multipole moment and consequently to the PN waveform of the GW signals.
We show that the magnetic field of the highly magnetized system can directly appear, at least, in the first PN order, hereafter 1\tiny PN \normalsize, of the quadrupole and 4-pole moments. We apply these PN quadrupole and 4-pole moments to obtain GWs generated by a single rotating magnetar. Here, we investigate the imprint of the magnetic field on GW signals up to the 1\tiny PN \normalsize order. We leave the higher-order contributions of the magnetic field, due to the multipole moments and the wave-zone part, for future works.

It should be noted that these calculations are the first analytic computation of the direct effect of the magnetic field on GW signals. 
We analytically display that the magnetic field of the fast spinning magnetar can be a potentially interesting source of GWs in the early stages of its life. We estimate that this type of GWs may be strong enough to be detected by the next generation of detectors, albeit with a low event rate.
 In fact, as long as the magnetic and rotation axes are not parallel, the magnetic field of this system itself can be a GW emitter.
On the other hand, because of the slow rotation of the 23 well-known magnetars, $T\sim 10\,\text{s}$, and the weak amplitude of GWs, of order $\sim 10^{-36}$, these systems do not seem to be the suitable source for generating GW signals by virtue of the presence of the magnetic field.

This paper is organized as follows. Sec. \ref{General review} is assigned to a brief review. In this section, we specify the strategy of our calculations. In Sec. \ref{rotating Magnetar}, by utilizing the simple and realistic models, we display how the magnetic field of the rotating magnetar can emit GW. Next, we investigate the characteristic strain of the GW emissions generated by the pure B-field part of the possible highly magnetized astrophysical sources. Furthermore, we compare it with the sensitivity range of the current and future GW detectors in Sec. \ref{Detectors}.
Finally, in Sec. \ref{Summary and conclusion}, we summarize the main results and conclude about them. Furthermore, in order to avoid displaying too many equations, we present most of the calculations in Appx. \ref{AppI}.

Throughout this paper, the Greek indices vary from 0 to 3. Also, the Latin indices stand for the coordinates $x$, $y$, and $z$. And $\eta^{\alpha \beta}=\text{diag}(-1,1,1,1)$ is the Minkowskian metric. Moreover, in the PN context, each order $c^{-2}$ represents the 1\tiny PN \normalsize correction.

\section{General review: the strategy of calculations}\label{General review}

In this section, for the sake of convenience, we present a short review and assemble the essential relations needed to study GWs in the presence of the magnetic field.
Let us start this short review by recalling the main ingredients of PN gravity. In this context, the velocity of the system elements is slow compared to the velocity of light and the corresponding gravitational potential field is weak. Furthermore, the calculations are restricted to the near zone where the field point $\bm{x}$ is situated within the three-dimensional sphere with radius $\mathcal{R}$, i.e., $\bm{x}<\mathcal{R}$. Beyond this region called the wave zone, i.e., $\bm{x}>\mathcal{R}$, post-Minkowskian theory (PM) is established.
In the modern approach of PN approximation, it is shown that PN theory is embedded within PM theory.

In Subsec. \ref{Post-Newtonian charged fluid}, we briefly review the PN version of the magnetohydrodynamics which has been comprehensively studied in \cite{nazari2018post}.
We first introduce the near-zone metric of the spacetime within a PN charged fluid and the associated PN gravitational potentials to 1\tiny PN \normalsize order. Then we obtain the effective energy-momentum pseudo tensor in which the B-field contribution is included. As illustrated in the next subsection, the components of this pseudo tensor construct the foundation of the PN gravitational potential, and consequently are helpful to investigate the GW propagation.

In Subsec. \ref{Gravitational waves}, we recall the PN expansion of the gravitational potential $h^{jk}$ in terms of its ingredients, i.e., the multipole moments, in PN gravity (\citealp{poisson2014gravity}). By considering the components of the energy-momentum tensor, we estimate the PN order of each multipole moment, and finally obtain the leading PN order in which the magnetic field can have a contribution. We then introduce the quadrupole formula that contains the B-field contribution.

\subsection{Post-Newtonian charged fluid}\label{Post-Newtonian charged fluid}

We begin in this subsection by reviewing the metric of a weakly curved spacetime within the charged fluid in the context of PN theory. In \cite{nazari2018post}, we obtained the near-zone metric in the PN limit where the characteristic velocity of system ingredients, $v_{\text{c}}$, is dramatically smaller than the velocity of light 
and the relevant gravitational fields are weak. 
We took advantage of the modern approach to PN theory where the harmonic gauge condition is established, and constructed the PN metric from its bases, i.e., the gravitational potentials $h^{\alpha\beta}$. The final result for the PN metric including the magnetic field's effect is 
\begin{subequations}
\begin{align}
\label{g00_2}
& g_{00}= -1+\frac{2}{c^2}U+\frac{2}{c^4}\bigg[\psi+\frac{1}{2}\partial_{tt}X+2\mathcal{B}-U^2\bigg]+O(c^{-6}),\\
\label{g0j_2}
& g_{0j} = -\frac{4}{c^3}U^j+O(c^{-5}),\\
\label{gjk_2}
& g_{jk} = \delta^{jk}\bigg(1+\frac{2}{c^2}U\bigg)+O(c^{-4}),\\
\label{det-g2}
&\left(-g\right)= 1+\frac{4}{c^2}U+O(c^{-4}),
\end{align}
\end{subequations}
where $U$ is the Newtonian gravitational potential and PN potentials $\psi$, $X$, $\mathcal{B}$, and $U^j$ are defined as 
\begin{subequations}
\begin{align}
\label{psi_int}
& \psi\left(t,\bm{x}\right)= G\int_{\mathcal{M}}\frac{{\rho^*}'\left(\frac{3}{2}v'^2-U'+\Pi'\right)+3p'}{|\bm{x}-\bm{x}'|}d^3x',\\
\label{X_int}
& X\left(t,\bm{x}\right)= G\int_{\mathcal{M}}{\rho^*}'|\bm{x}-\bm{x}'|d^3x',\\
\label{B_int}
& \mathcal{B}\left(t,\bm{x}\right)= G\int_{\mathcal{M}}\frac{1}{2\,\mu_0}\frac{B'^2}{|\bm{x}-\bm{x}'|}d^3x',\\
\label{U^J_int}
& U^j\left(t,\bm{x}\right)= G\int_{\mathcal{M}}\frac{{\rho^*}'v'^j}{|\bm{x}-\bm{x}'|}d^3x',
\end{align}
\end{subequations}
respectively. In these integrations, $\bm{v}$ is fluid's velocity field, $\bm{B}$ is the magnetic field, $\Pi$ is the ratio of the proper internal energy density $\epsilon$ to the rescaled mass density $\rho^*$ given by
\begin{align}
\rho^*=\rho\bigg[1+\frac{1}{2}\left(\frac{v}{c}\right)^2\bigg]+O(c^{-4}),
\end{align}
and $p$ is the thermal pressure.  
For more detailed discussions on the derivation of this metric in the PN limit, we refer the reader to section 2 of \cite{nazari2018post}.
Here, the primed variables are considered at time $t$ and  position $\bm{x}'$. Also, the domain of
the integrates, $\mathcal{M}$, is a sphere with an arbitrary radius $\mathcal{R}$ which is $\mathcal{R}<\lambda_{\text{c}}$.  $\lambda_{\text{c}}$ is the characteristic wavelength of GWs generated by the source.

In the modern approach of PN theory, the gravitational potentials $h^{\alpha\beta}$ are constructed from the effective energy-momentum pseudotensor $\tau^{\alpha\beta}=\left(-g\right)(T^{\alpha\beta}+t_{\text{LL}}^{\alpha\beta}+t_{\text{H}}^{\alpha\beta})$ (\citealp{poisson2014gravity}).
As we shall see in the following, for calculating the contribution of electromagnetic fields to the emission of GW, it is necessary to have enough information about $\tau^{\alpha\beta}$. As shown in \cite{poisson2014gravity}, this effective pseudotensor is constructed from the energy-momentum tensor, $T^{\alpha\beta}$, the Landau-Lifshitz, and harmonic-gauge pseudotensors, i.e., $(-g)t_{\text{LL}}^{\alpha\beta}$, and $(-g)t_{\text{H}}^{\alpha\beta}$, respectively. In \cite{nazari2018post}, we showed that the electromagnetic fields do not appear in the latter pseudotensors to the 1\tiny PN \normalsize approximation. So, we drop these terms and build the effective energy-momentum pseudotensor from the rest of its ingredients, i.e., from $\left(-g \right)T^{\alpha\beta}$.

Because of the presence of electromagnetic fields in this problem, $T^{\alpha\beta}$ includes both the matter and B-field contributions, $T^{\alpha\beta}=T^{\alpha\beta}_{\text{fluid}}+T^{\alpha\beta}_{\text{field}}$. For the matter portion, we assume that fluid is perfect. So we have 
\begin{align}\label{T_fluid}
T^{\alpha\beta}_{\text{fluid}}=\left(\rho+\frac{\epsilon}{c^2}+\frac{p}{c^2}\right)u^{\alpha}u^{\beta}+pg^{\alpha\beta},
\end{align} 
where $u^{\alpha}=\gamma\left(c,\bm{v}\right)$ is the velocity field in which $\gamma=u^0/c$. Furthermore, the contribution of electromagnetic fields to the energy-momentum tensor in terms of the electromagnetic field tensor $F^{\alpha\beta}$, is expressed as
\begin{equation}\label{T_field}
T^{\alpha\beta}_{\text{field}}=\frac{1}{\mu_0}\left(F^{\alpha\mu}F^{\beta}_{\mu}-\frac{1}{4}g^{\alpha\beta}F_{\mu\nu}F^{\mu\nu}\right),
\end{equation}
where  $\mu_0$ is the vacuum permeability. The components of $F^{\alpha\beta}$ are given by $F^{0j}=c^{-1}E^j$ and $F^{ij}=\epsilon^{ijk}B_{k}$. Here, $\epsilon^{ijk}$ is the permutation symbol, and the electric $\bm{E}$ and magnetic $\bm{B}$ fields are measured in the laboratory framework. 
By utilizing the components of the metric  \eqref{g00_2}-\eqref{gjk_2}, we obtain the covariant components of the electromagnetic field tensor 
\begin{subequations}
\begin{align}
\label{covar_F0i}
& F_{0i}=-\frac{E^i}{c}+\frac{4}{c^3}\left(\bm{U}\times\bm{B}\right)^i+O(c^{-5}),\\
\label{covar_Fij}
& F_{ij}=\epsilon_{ijk}B^k\left(1+\frac{4U}{c^2}\right)+O(c^{-4}),
\end{align}
\end{subequations}
and consequently find the energy-momentum tensor as 
\begin{subequations}
\begin{align}
\label{T00_2}
& T^{00}= c^2\rho^*\left[1+\frac{1}{c^2}\Big(\frac{v^2}{2}+\Pi-U\Big)\right]+\frac{B^2}{2\mu_0}+O(c^{-2}),\\
\nonumber
& T^{0j}= c\rho^*v^j\left[1+\frac{1}{c^2}\Big(\frac{v^2}{2}+\Pi-U+\frac{p}{\rho^*}\Big)\right]\\\label{T0j_2}
&~~~~~~~~~~~~~~~~~~~~~~~~~~~~~~~~~~~~~~~~~+\frac{1}{\mu_0 c}\left(\bm{E}\times\bm{B}\right)^j+O(c^{-3}),\\
\nonumber
& T^{jk}= \rho^*v^jv^k\left[1+\frac{1}{c^2}\Big(\frac{v^2}{2}+\Pi-U+\frac{p}{\rho^*}\Big)\right]+p\Big(1-\frac{2U}{c^2}\Big)\delta^{jk}\\\nonumber
&-\frac{1}{\mu_0c^2}\left[E^jE^k-\frac{E^2}{2}\delta^{jk}\right]-\frac{1}{\mu_0}\left[B^jB^k-\frac{B^2}{2}\delta^{jk}\right]\Big(1+\frac{2U}{c^2}\Big)\\\label{Tjk_2}
&+O(c^{-4}),
\end{align}
\end{subequations}
after combining $T^{\alpha\beta}_{\text{fluid}}$ and $T^{\alpha\beta}_{\text{field}}$. 
Since $T^{\alpha\beta}_{\text{field}}$ is traceless, then there are magnetic contributions in $T^{00}$ and $T^{jk}$ with the same PN orders.
One can easily obtain the effective energy-momentum pseudotensor by multiplying above components by Eq. \eqref{det-g2}.

\subsection{Gravitational waves}\label{Gravitational waves}

The main task to find the GW field is to obtain the space-space component of the gravitational potentials $h^{\alpha\beta}$ in the far-away wave zone where the radiative aspects are important. \footnote{ In chapter 6 and 7 of \cite{poisson2014gravity}, it is comprehensively described that by applying the Landau-Lifshitz formulation of the Einstein field equations, known as the modern approach to PN theory, how one can obtain $h^{\alpha\beta}$ in the near and wave zones. }

In the PN context, the gravitational potentials $h^{jk}$ in the far-away wave zone are given by
\begin{align}\label{hjkT}
& h^{jk}=\frac{2G}{c^4 R_{\text{d}}}\frac{\partial^2}{\partial \tau^2}\bigg\lbrace Q^{jk}+Q^{jka}N_{a}+Q^{jkab}N_a N_b\\\nonumber
&~~~~~~+\frac{1}{3}Q^{jkabc}N_a N_b N_c+\cdots\bigg\rbrace+\frac{2G}{c^4 R_{\text{d}}}\bigg\lbrace P^{jk}+P^{jka}N_a\bigg\rbrace,
\end{align}
in which   
$Q^{jk}$, $Q^{jka}$, $Q^{jkab}$, and $Q^{jkabc}$ are the multipole moments, and $P^{jk}$ and $P^{jka}$ are surface integrations given by Eqs (11.119a)-(11.120b) in \cite{poisson2014gravity}, respectively.
In the above relation, $\bm{N}=\bm{x}/R_{\text{d}}$ is a unit vector in the $\bm{x}$ direction, $R_{\text{d}}=\arrowvert \bm{x}\arrowvert$ is the distance to the field point $\bm{x}$, and $\tau=t-R_{\text{d}}/c$ is the retarded time.
Moreover, the multipole moments and surface integrations are functions of $\tau$ only, and the source of them, i.e., $\tau^{\alpha\beta}$ is a function of $\tau$ and $\bm{x}'$ where $\bm{x}'<\mathcal{R}$. And $h^{jk}$ is a function of $\tau$ and $\bm{x}$ located in the wave zone, $\bm{x}>\mathcal{R}$.

As mentioned previously, the goal of this paper is to survey the direct contribution of the electromagnetic fields to GW.  In the following, we estimate the order of magnitude of the multipole moments and surface integrals, and specify the terms containing electromagnetic effects that appear in the gravitational potential.
By considering Eqs. \eqref{T00_2}-\eqref{Tjk_2} and the definition of the multipole moments, we can symbolically indicate the $c^n$ orders as $Q^{jk}= c^{0}A_1+c^{-2}A_2+\cdots$ in which the magnetic field has a portion in order $c^{-2}$, $Q^{jka}=c^{-1}B_1+c^{-3}B_2+\cdots$ in which the electric and magnetic fields both contribute at order $c^{-3}$, $Q^{jkab}=c^{-2}C_1+c^{-4}C_2+\cdots$ in which the magnetic field has portions in orders $c^{-2}$ and $c^{-4}$ and the electric field has only a portion of order $c^{-4}$, and $Q^{jkabc}=c^{-3}D_1+c^{-5}D_2+\cdots$ in which the magnetic field has portions in orders $c^{-3}$ and $c^{-5}$ and the electric field contribution is of order $c^{-5}$.
For the surface integrals, by taking into account Eq. \eqref{Tjk_2}, we see that $P^{jk}=c^{0}E_1+c^{-2}E_2+\cdots$ where the magnetic field has portions in orders $c^{0}$ and $c^{-2}$ and $P^{jka}=c^{-1}F_1+c^{-3}F_2+\cdots$ where the magnetic field contributes at orders $c^{-1}$ and $c^{-3}$.

Here, as a good approximation, we assume that $\mathcal{M}$ is sufficiently large so that it contains the matter distribution and its associated electromagnetic fields. Therefore, the matter distribution, as well as the electromagnetic fields, can be ignored on its boundary, $\partial\mathcal{M}$.
Consequently, the magnetic field contributions due to the surface integrals $P^{jk}$ and $P^{jka}$ evaluated on $\partial\mathcal{M}$, which are of orders 0\tiny PN\normalsize, 0.5\tiny PN\normalsize, 1\tiny PN\normalsize, and 1.5\tiny PN\normalsize,  may be automatically omitted from the gravitational potential. 
Keeping this fact in mind and considering the estimations listed above, we deduce that the magnetic field plays a role at least in 1\tiny PN \normalsize order in the multipole moments $Q^{jk}$ and $Q^{jkab}$. 
In the present work, as a preliminary evaluation of the direct influence of the magnetic field on the GW fields, we only endeavor to compute the dominant terms in the definition of $Q^{jk}$  and $Q^{jkab}$.

It should be added that the source of Eq. \eqref{hjkT}, $\tau^{\alpha\beta}$, is a function of $\tau$ and $\bm{x}'$ where $\bm{x}'<\mathcal{R}$. This relation in fact represents the near-zone portion of the total potentials $h^{jk}=h^{jk}_{\mathcal{N}}+h^{jk}_{\mathcal{W}}$ which is evaluated in the wave zone. Therefore, to compute the potentials, one should obtain the wave-zone contribution to $h^{jk}$. In this case, both the field and source points lie within the wave zone, i.e.,  $\bm{x}>\mathcal{R}$ and $\bm{x}'>\mathcal{R}$. Here, by a simple estimation, we clarify that the electromagnetic field contribution to $h^{jk}_{\mathcal{W}}$ can be at least of 2.5 \tiny PN \normalsize order and consequently does not play a role in the required 1 \tiny PN \normalsize correction. As we know, in addition to the sources situated in the near zone, the wave-zone field energy can emit GW. This part of the GW field is known as the tails. It is shown that for the typical systems, i.e., when the relevant system has no electromagnetic fields, the energy of the matter field constructs the wave-zone portion of GW signals and consequently the tails are of the order of 1.5\tiny PN \normalsize (the tails and their PN order are comprehensively obtained in Sec. 11.3.7 of \cite{poisson2014gravity}.) On the other hand, if we choose a system associated with an electromagnetic field, its energy will be at least of the order $O(c^{-2})$ compared to the matter field energy. Therefore, one can roughly estimate that electromagnetic field portion of $h^{jk}_{\mathcal{W}}$ is a term of $c^{-3}\times c^{-2}$ order (2.5 \tiny PN \normalsize correction). Accurate computation of this correction and its PN order require long and tedious calculations that are beyond the scope of this paper. So, here, we benefit this estimation and leave more detailed calculations for a future work.

Therefore, by considering the above-mentioned points, Eq. \eqref{hjkT} reduces to
\begin{equation}\label{hjk}
h^{jk}=\frac{2 G}{c^4 R_{\text{d}}}\big(\ddot{Q}^{jk}+\ddot{Q}^{jkab}N_aN_b\big),
\end{equation}
in which overdots represent the second derivative with respect to the retarded time $\tau$ and
\begin{subequations}
\begin{align}
\nonumber
& Q^{jk}=\int_{\mathcal{M}}\bigg[\rho^*+\frac{1}{c^2}\Big(\rho^*\big(\frac{1}{2}v^2+\Pi+3U\big)+\frac{B^2}{2\mu_0}\Big)\bigg]x^jx^kd^3x\\\label{QjkR}
&-\frac{7}{8\pi G c^2}\int_{\mathcal{M}}\partial_m U \partial^m U x^jx^kd^3x+O(c^{-4}),\\
\label{QjkabR}
&Q^{jkab}=\\\nonumber
&\frac{1}{c^2}\int_{\mathcal{M}}\bigg[\rho^*v^jv^k+p\delta^{jk}-\frac{1}{\mu_0}\Big(B^jB^k-\frac{1}{2}B^2\delta^{jk}\Big)\bigg]x^ax^bd^3x\\\nonumber
&+\frac{1}{4\pi G c^2}\int_{\mathcal{M}}\Big(\partial^jU\partial^kU-\frac{1}{2}\delta^{jk}\partial_mU\partial^mU\Big)x^ax^bd^3x+O(c^{-4}),
\end{align}
\end{subequations}
can be obtained by inserting Eqs. \eqref{det-g2} and \eqref{T00_2} within the definition of $Q^{jk}$ and Eq. \eqref{Tjk_2} within the definition of $Q^{jkab}$, respectively. 
For simplicity, we dropped the prime marks in the integrands.
The second integrals in Eqs. \eqref{QjkR} and \eqref{QjkabR} come from the Landau-Lifshitz and harmonic-gauge pseudotensors (\citealp{poisson2014gravity}). It should be recalled that electromagnetic fields do not emerge in the Landau-Lifshitz and harmonic contributions at 1\tiny PN \normalsize order.
In the following, for simplicity and to clearly reveal the significant role of the magnetic field, we ignore the other 1\tiny PN \normalsize terms in the integrals \eqref{QjkR} and \eqref{QjkabR} that come from the matter portion and only compare 1\tiny PN \normalsize B-field signals with the leading material contribution. So, we finally arrive at 
\begin{subequations}
\begin{align}
\label{Ijk}
&\mathcal{I}^{jk}=\mathcal{I}^{jk}_{\text{M}}+\mathcal{I}^{jk}_{\text{F}},\\
\label{Qjka}
&\mathcal{Q}^{jkab}=\mathcal{Q}^{jkab}_{\text{F}}+\delta^{jk}\mathcal{I}^{ab}_{\text{F}},
\end{align}
\end{subequations}
where
\begin{subequations}
\begin{align}
\label{IjkM}
&\mathcal{I}^{jk}_{\text{M}}\left(\tau\right)= \int_{\mathcal{M}}\rho^*\left(\tau,\bm{x}\right)x^jx^kd^3x,\\
\label{IjkF}
&\mathcal{I}^{jk}_{\text{F}}\left(\tau\right)= \frac{1}{2\mu_0c^2}\int_{\mathcal{M}}B^{2}\left(\tau,\bm{x}\right)x^jx^kd^3x,\\
\label{QFjkab}
&\mathcal{Q}^{jkab}_{\text{F}}\left(\tau\right)=-\frac{1}{\mu_0c^2}\int_{\mathcal{M}}B^j\left(\tau,\bm{x}\right)B^k\left(\tau,\bm{x}\right)x^ax^bd^3x.\color{black}
\end{align}
\end{subequations} 
Eq. \eqref{Ijk} exhibits the leading order only including the matter and B-field contributions, i.e.,  $\mathcal{I}^{jk}_{\text{M}}$ and $\mathcal{I}^{jk}_{\text{F}}$ respectively, to the PN expansion of $Q^{jk}$. Similarly, Eq. \eqref{Qjka} represents the B-field portion of the PN expansion of $Q^{jkab}$. Here, we choose the notations $\mathcal{I}^{jk}$, known as the mass quadrupole-moment tensor, and $\mathcal{Q}^{jkab}$ for these leading contributions. In fact, we carry out our calculation beyond the famous quadrupole formula $h^{jk}_{\text{M}}=\left(2 G/c^4R_{\text{d}}\right)\ddot{\mathcal{I}}^{jk}_{\text{M}}$ by incorporating $\mathcal{I}^{jk}_{\text{F}}$ and $\mathcal{Q}^{jkab}$ into this relation.

It should be stressed that some part of $\mathcal{Q}^{jkab}$ which is proportional to $\delta^{jk}$, has no contribution to the transverse-tracefree piece of $h^{jk}$ and consequently to the GW field. So, hereafter, we drop this term and reduce Eq. \eqref{Qjka} to $\mathcal{Q}^{jkab}=\mathcal{Q}^{jkab}_{\text{F}}$.
Therefore, in summary, to find the direct effect of the magnetic field on GWs, the transverse-tracefree piece of $h^{jk}_{\text{F}}=\left(2 G/c^4R_{\text{d}}\right)\big(\ddot{\mathcal{I}}^{jk}_{\text{F}}+\ddot{\mathcal{Q}}^{jkab}_{\text{F}}N_aN_b\big)$ must be calculated.
In Sec. \ref{rotating Magnetar}, we attempt to obtain this part of the GW field for a single rotating magnetar.

\section{Rotating Magnetar}\label{rotating Magnetar}

As a possible source of the GW emission, we investigate the rotating NS. 
Since our goal is to study the contribution of the magnetic field to GW signals, we assume that NS is highly magnetized and its B-field magnitude is about $\sim 10^{15}\,\text{G}$. As we know, this type of NS is called magnetar.
In Subsec. \ref{Simple model}, by introducing a simple model, the vacuum rotating dipole model, we attempt to describe the B-field configuration inside and outside NS. Utilizing this model, we display how GW can change in the presence of such B-field shape.
In Subsec. \ref{Discussion}, we evaluate the behavior of GWs generated by the B-field part of the rotating magnetar and compare this portion with those of the matter part.
Then we compare the results obtained from this simple model with a more realistic model where the magnetar has a corotating dense magnetosphere in Subsec. \ref{Realistic model}. 
This comparison reveals that the results of the simple model are reliable and can be used to interpret GWs emitted from the magnetic field of the real magnetar up to 1\tiny PN \normalsize order.
Finally, in Subsec. \ref{Application to magnetars}, applying the results of this section, we study the millisecond and 21 well-known magnetars\footnote{As we know, so far the 23 candidates have been introduced for the magnetars. But, as reported in \cite{olausen2014mcgill}, we do not have enough information to calculate the GW signals form two cases of these systems.} and obtain the amplitude of GW signals which can be produced by these systems.

\subsection{Simple model}\label{Simple model}

In this subsection, we launch our calculation by choosing a normal model simply describing the configuration of the magnetic field inside and outside of magnetars.
Since our goal is to study analytically the role of the magnetic field in GW signals, it makes sense to choose the simplest B-field configuration as the first step in this route.
We consider that in the absence of rotation, a star is a uniformly magnetized sphere with a constant magnetization $\bm{\mu}_{\text{m}}$ given by
 \begin{equation}
\bm{\mu}_{\text{m}}=\mu_{\text{m}} \bm{e}_r,
\end{equation}
in which $\mu_{\text{m}}$ is a constant parameter and $\bm{e}_r$ is a constant unit vector indicating the direction of $\bm{\mu}_{\text{m}}$. For the magnetic dipole moment, we have $\bm{m}=\left(4\pi/3\right) a^3 \bm{\mu}_{\text{m}}$ where $a$ is the radius of the star. 
This magnetized star produces the uniform magnetic field inside and the dipole form outside the star (\citealp{jackson2007classical}). So we have
\begin{subequations}
\begin{align}
\label{Bin}
&\bm{B}_{\text{in}}=\frac{2}{3}\mu_0\bm{\mu}_{\text{m}}~~~~~~~~~~~~~~~~~~~~~~~~~~~~~~~~~~~~~~~~~~\text{for}~~r<a,\\
\label{Bout}
&\bm{B}_{\text{out}}(\bm{x})=\frac{\mu_0}{4\pi}\frac{3\bm{n}\left(\bm{m}\cdot\bm{n}\right)-\bm{m}}{r^3}~~~~~~~~~~~~~~\text{for}~~r>a,
\end{align}
\end{subequations} 
where $\bm{n}=\bm{x}/r$ is the unit vector in the direction $\bm{x}$, and $r=\arrowvert \bm{x}\arrowvert$.
Also, we assume that this star rigidly rotates around one of its principal axes with the angular velocity $\Omega$. Here, without loss of generality, the $z$-axis is chosen to be parallel to the rotation axis. 
Here, we also assume that $\bm{\mu}_{\text{m}}$ is not necessarily parallel to $\bm{\Omega}$ and the angle between them remains constant.
Therefore, the moving magnetic axis, which is along the magnetization, creates a cone around the $z$-axis.  For simplification, there is no magnetosphere around the magnetar in this description. In fact, we consider a rotating dipole toy model in the vacuum for magnetars. Of course, in the following subsection, we will discuss a more realistic model.

It should be noted that this simple model of the B-field configuration is also applied in \cite{bonazzola1996gravitational}. It is shown that in the general case, because of the magnetic stresses, the Newtonian fluid star is deformed and has the triaxial ellipsoid shape that rotates around an axis being none of its principal axes. However, here, for the sake of simplification, we choose that $\bm{\Omega}$ of this deformed star coincides with one of its principal axes. Furthermore, in contrast to the case of the magnetically deformed star, we consider the general case in which in addition to the magnetic field, the distortion of the star can be induced by other physical phenomena. In the following subsections, we compare our results with both this general case and the magnetically deformed star.

In order to find GWs emitted by the magnetic field of this system, as a first task, we should derive the B-field contribution of the quadrupole and  4-pole moments, i.e., $\mathcal{I}^{jk}_{\text{F}}$ and $\mathcal{Q}^{jkab}_{\text{F}}$, respectively. We first find these tensors in the coordinate frame that rotates with the star. We indicate this corotating frame with $\left(x, y, z\right)$. Then, by using the appropriate coordinate transformation, we obtain the corresponding tensors in the non-rotating frame which is shown by $\left(x', y', z'\right)$. 
 To do so, we rewrite the magnetization $\bm{\mu}_{\text{m}}$ in the Cartesian coordinate system as
\begin{align}
\label{mu}
\bm{\mu}_{\text{m}}=\mu_{\text{m}}\left(\sin\theta_0\cos\varphi_0\bm{e}_x+\sin\theta_0\sin\varphi_0\bm{e}_y+\cos\theta_0\bm{e}_z \right),
\end{align} 
where $\theta_0$ and $\varphi_0$ are the angles between the magnetization and $z$ and $x$-axes in the rotating frame, respectively. 
Then, by inserting the above relation within Eqs. \eqref{Bin} and \eqref{Bout}, we obtain $B_{\text{in}}^{2}$ and $B_{\text{out}}^{2}$ as
\begin{subequations}
\begin{align}
\label{Bin cm}
& B_{\text{in}}^{2}=\frac{4}{9}\mu_0^2 \mu_{\text{m}}^2,\\
\nonumber
& B_{\text{out}}^{2}=\frac{a^6}{9r^6}\mu_0^2\mu_{\text{m}}^2\bigg[3\Big(\cos\theta\cos\theta_0\\\label{Bout cm}
&~~~~~~~~~~~~~~~~~~~~~~~~~~~~~+\cos\left(\varphi-\varphi_0\right)\sin\theta\sin\theta_0\Big)^2+1\bigg],
\end{align}
\end{subequations}
respectively. Where $\theta$ and $\varphi$ uniquely determine the direction of the unit vector $\bm{n}$. 
We are now in a position to obtain Eq. \eqref{IjkF} in the rotating frame.  
To carry out the integration over sphere $\mathcal{M}$, we separate the domain of the integral into two regions, the interior and exterior of the star. So, for Eq. \eqref{IjkF}, we have  
\begin{align}
\label{IF}
&\mathcal{I}^{jk}_\text{F}\left(\tau\right)=\frac{1}{2\mu_0c^2}\int_{\mathcal{M}}B^{2}\left(\tau,\bm{x}\right)x^jx^kd^3x\\\nonumber
&=\frac{1}{2\mu_0c^2}\left[\int_{r<a} B^{2}_{\text{in}}\left(\tau,\bm{x}\right)x^jx^kd^3x
+\int_{r>a} B^{2}_{\text{out}}\left(\tau,\bm{x}\right)x^jx^kd^3x\right].
\end{align}

In the following, by substituting Eqs. \eqref{Bin cm} and \eqref{Bout cm} into the above relation, we derive the $x$-$x$ component of $\mathcal{I}^{jk}_\text{F}$. For other components, the same method as the one introduced here will be applied. We obtain $\mathcal{I}^{xx}_{\text{F}}$ as 
\begin{align}
\nonumber
&\mathcal{I}^{xx}_{\text{F}}=\frac{1}{2\mu_0 c^2}\left[\int_{r<a}B^{2}_{\text{in}}\, x^2 d^3x+\int_{r>a} B^{2}_{\text{out}}\,x^2 d^3x\right]\\\nonumber
& = \frac{1}{2\mu_0 c^2}\bigg[\int f_{\text{in}}\left(\theta,\varphi\right) d\Omega\int_0^a  r^4  d r \\
&~~~~~~~~~~~~~~~~~~~~~~~~~~~~~~~~~~~~~~+\int f_{\text{out}}\left(\theta,\varphi\right) d\Omega\int_a^{\mathcal{R}}  r^{-2}  d r \bigg],
\end{align}
where  $f_{\text{in}}\big(\theta, \varphi\big) = B^{2}_{\text{in}} \sin^2\theta \cos^2\varphi$,  $f_{\text{out}}\big(\theta, \varphi\big) = r^6 B^{2}_{\text{out}} \sin^2\theta \cos^2\varphi$, and $d\Omega = \sin \theta d\theta d\varphi$ is an element of the solid angle. As seen, the calculation of these integrals is straightforward. The only remarkable point in this calculation is that, after deriving the radial integral related to the star exterior, we can freely discard $\mathcal{R}$-dependent terms. In fact, in the modern approach of PN approximation, it is shown that the $\mathcal{R}$-dependent terms of the wave-zone contribution cancel the corresponding terms arising from the near-zone part. So, one can remove these terms from the final answer. This cancellation
is justifiably introduced in chapters 6 and 7 of \cite{poisson2014gravity}. 
By using this method and consequently dropping the $\mathcal{R}$-dependent term, we finally arrive at
\begin{align}
\label{Ixx}
\mathcal{I}^{xx}_{\text{F}}=A_0\big(9-\cos 2\theta_0+2\sin^2\theta_0\cos 2\varphi_0 \big),
\end{align}
where $A_0=\frac{1}{c^2}\frac{\pi}{45}\mu_0 \mu_{\text{m}}^2 a^5$. The other components of $\mathcal{I}^{jk}_{\text{F}}$ can be similarly calculated as follows
\begin{subequations}
\begin{align}
\label{Iyy}
&\mathcal{I}^{yy}_{\text{F}}= A_0\left(9-\cos 2\theta_0-2\sin^2\theta_0\cos 2\varphi_0 \right),\\
\label{Izz}
&\mathcal{I}^{zz}_{\text{F}}= 2A_0\left(5+\cos 2\theta_0 \right),\\
\label{Ixy}
&\mathcal{I}^{xy}_{\text{F}}= 2A_0 \sin^2\theta_0\sin 2\varphi_0\,,\\
\label{Ixz}
&\mathcal{I}^{xz}_{\text{F}}= 2A_0 \sin 2\theta_0\cos\varphi_0\,,\\
\label{Iyz}
&\mathcal{I}^{yz}_{\text{F}}= 2A_0 \sin 2\theta_0\sin\varphi_0\,.
\end{align}
\end{subequations}
To obtain the components of the 4-pole moment $\mathcal{Q}^{jkab}_{\text{F}}$, we utilize the same method as mentioned above. We present the results in Appx. \ref{AppI}.

One can see that, as expected, these components do not depend upon the time. Therefore, if this system does not rotate, GW cannot be generated by a static and uniformly magnetized sphere.
Now, we assume that this sphere rigidly rotates. So, by utilizing the transformation  $\left(x, y, z\right)\longrightarrow\left(x', y', z'\right)$ which is given by
\begin{subequations}
\begin{align}
x'&= x\cos\Omega t-y\sin\Omega t,\\
y'&= x\sin\Omega t + y\cos\Omega t,\\
z'&= z, 
\end{align}
\end{subequations}
and then applying the tensor transformation law 
\begin{align}
\label{moment tensor}
&\mathcal{I}^{a'b'}=\frac{\partial x^{a'}}{\partial x^j}\frac{\partial x^{b'}}{\partial x^k}\mathcal{I}^{jk},\\
\label{moment-tensor1}
& \mathcal{Q}^{j'k'l'm'}=\frac{\partial x^{j'}}{\partial x^a}\frac{\partial x^{k'}}{\partial x^b}\frac{\partial x^{l'}}{\partial x^c}\frac{\partial x^{m'}}{\partial x^d}\mathcal{Q}^{abcd},
\end{align}
we can find the components of $\mathcal{I}^{jk}_{\text{F}}$ and $\mathcal{Q}^{jkab}_{\text{F}}$ in the non-rotating frame. For  $\mathcal{I}^{j'k'}_{\text{F}}$, we have
\begin{subequations}
\begin{align}
\label{IxxF}
&\mathcal{I}^{x' x'}_{\text{F}}= A_0\left(9-\cos 2\theta_0+2\sin^2\theta_0\cos 2(\Omega t+\varphi_0)\right),\\
&\mathcal{I}^{y' y'}_{\text{F}}= A_0\left(9-\cos 2\theta_0-2\sin^2\theta_0\cos 2(\Omega t+\varphi_0) \right),\\
&\mathcal{I}^{z' z'}_{\text{F}}= 2A_0\left(5+\cos 2\theta_0 \right),\\
&\mathcal{I}^{x' y'}_{\text{F}}= 2A_0 \sin^2\theta_0\sin 2\left(\Omega t +\varphi_0\right),\\
&\mathcal{I}^{x' z'}_{\text{F}}= 2A_0 \sin 2\theta_0\cos \left(\Omega t +\varphi_0\right),\\
\label{IyzF}
&\mathcal{I}^{y' z'}_{\text{F}}= 2A_0 \sin 2\theta_0\sin \left(\Omega t +\varphi_0\right).
\end{align}
\end{subequations}
The components of the 4-pole-moment tensor in the non-rotating frame are given by Eqs. \eqref{Qx'x'x'x'}-\eqref{Qz'z'x'x'}.

According to Eq \eqref{hjk}, the next task is to evaluate the second time derivatives of $\mathcal{I}^{j'k'}_{\text{F}}$ and $\mathcal{Q}^{j'k'a'b'}_{\text{F}}$. However, let us first introduce the matter part of the quadrupole moment, i.e., $\mathcal{I}^{jk}_{\text{M}}$ and complete the total form of $\mathcal{I}^{jk}$. 
This portion is completely derived in chapter 11 of \cite{poisson2014gravity}. We very briefly review these calculations here. As shown in this reference, for a deformed rotating star, the non-vanishing components of the quadrupole-moment tensor $\mathcal{I}^{jk}_{\text{M}}$ are given by 
\begin{subequations}
\begin{align}
\label{IxxM}
\mathcal{I}_{\text{M}}^{x'x'}&=\frac{1}{2}I_3-\frac{1}{2}\left(I_1-I_2\right)\cos 2\Omega t,\\
\mathcal{I}_{\text{M}}^{y'y'}&=\frac{1}{2}I_3+\frac{1}{2}\left(I_1-I_2\right)\cos 2\Omega t,\\
\mathcal{I}_{\text{M}}^{z'z'}&=\frac{1}{2}\left(I_1+I_2-I_3\right),\\
\label{IxyM}
\mathcal{I}_{\text{M}}^{x'y'}&=-\frac{1}{2}\left(I_1-I_2\right)\sin 2\Omega t,
\end{align}
\end{subequations} 
in the non-rotating frame. In the above equations, $I_1$, $I_2$, and $I_3$ are the principle moments of inertia which are respectively illustrated as
\begin{subequations}
\begin{align}
I_1&=\int_{\mathcal{M}}\rho^*(\bm{x})\left(y^2+z^2\right)d^3x=\mathcal{I}_{\text{M}}^{yy}+\mathcal{I}_{\text{M}}^{zz},\\
I_2&=\int_{\mathcal{M}}\rho^*(\bm{x})\left(x^2+z^2\right)d^3x=\mathcal{I}_{\text{M}}^{xx}+\mathcal{I}_{\text{M}}^{zz},\\
I_3&=\int_{\mathcal{M}}\rho^*(\bm{x})\left(x^2+y^2\right)d^3x=\mathcal{I}_{\text{M}}^{xx}+\mathcal{I}_{\text{M}}^{yy},
\end{align}
\end{subequations}
in terms of $\mathcal{I}_{\text{M}}^{xx}$, $\mathcal{I}_{\text{M}}^{yy}$, and $\mathcal{I}_{\text{M}}^{zz}$ in the corotating frame. 
It is worth noting that in this case, it is assumed the $x$-axis shows the direction of the largest axis of the star. In fact, by this assumption, we consider that the magnetic field lines are not parallel with the deformation direction of the star, i.e., $\theta_0\neq \pi/2$ and $\varphi_0\neq 0$. For example, one can consider a system that strong factors such as the centrifugal force and internal thermal pressure and also external factors such as the tidal force from a nearby system may cause to deform in different directions from those of the magnetic field. So, by considering these effects, the above assumption is not far from reality.

Now, we return to find the GW field. To do so, we first integrate Eqs. \eqref{IxxF}-\eqref{IyzF} and \eqref{IxxM}-\eqref{IxyM} and construct the components of $\mathcal{I}^{a'b'}$. Then by evaluating the time derivatives of the total quadrupole-moment tensor $\mathcal{I}^{a'b'}$, after some manipulations, we obtain the nonzero components as
\begin{subequations}
\begin{align}
\label{IFxx}
&\ddot{\mathcal{I}}^{x'x'}= 2\epsilon_{\text{M}} I_3\Omega^2\cos 2\Omega t-8 A_0\Omega^2\sin^2\theta_0\cos 2\left(\Omega t +\varphi_0\right),\\
\label{IFyy}
&\ddot{\mathcal{I}}^{y'y'}=-2\epsilon_{\text{M}} I_3\Omega^2 \cos 2\Omega t+8 A_0\Omega^2\sin^2\theta_0\cos 2\left(\Omega t +\varphi_0\right),\\
\label{IFxy}
&\ddot{\mathcal{I}}^{x' y'}= 2\epsilon_{\text{M}} I_3\Omega^2\sin 2\Omega t-8 A_0 \Omega^2\sin^2\theta_0\sin 2\left(\Omega t +\varphi_0\right),\\
\label{IFxz}
&\ddot{\mathcal{I}}^{x' z'}= -2 A_0 \Omega^2\sin 2\theta_0\cos \left(\Omega t +\varphi_0\right),\\
\label{IFyz}
&\ddot{\mathcal{I}}^{y' z'}= -2 A_0 \Omega^2 \sin 2\theta_0\sin \left(\Omega t +\varphi_0\right),
\end{align}
\end{subequations}
in which $\epsilon_{\text{M}}=\left(I_1-I_2\right)/I_3$ is know as the ellipticity parameter measuring the deformation of the body.
In the next step, we obtain the components of $\ddot{\mathcal{Q}}^{j'k'a'b'}_{\text{F}}$. As this part of the calculation is very long, we have removed it from the main text of this paper and introduced these terms in Appx. \ref{AppI}.

Substituting Eqs. \eqref{IFxx}-\eqref{IFyz} and
\eqref{Qxxzz}-\eqref{Qyzzz} into Eq. \eqref{hjk}, after some manipulations, we obtain the components of the gravitational potential $h^{jk}$. Therefore, as a final task in this subsection, we should find the transverse-tracefree piece of this tensor, namely $h^{jk}_{\text{TT}}$, needed to build the GW field. To do so,
we extract two independent polarizations of the GW tensor $h^{jk}_{\text{TT}}$, i.e., $h_{+}$ and $h_{\times}$, which exist on the transverse subspace. In order to demonstrate this subspace, it is common to introduce a "detector-adapted" frame $\left(X, Y, Z\right)$. It is assumed that the origin of this frame coincides with the origin of non-rotating frame $\left(x', y', z'\right)$. Also, the $Z$-axis represents the direction of the GWs propagation toward the GW detector, and it is orthogonal to the transverse subspace $\left(X, Y\right)$. The angle between $Z$ and $z'$ axes is shown by $\iota$. This angle is called the inclination angle. Moreover, in the detector frame, the $X$-axis is the intersection of the plane of the sky with the equatorial plane of the star. As mentioned before, it is considered that, in the rotating coordinate system, the $x$-axis is parallel to the long axis of the deformed star. At $t=0$ when $x'=x$, the angle between $X$ and $x'$ axes is exhibited by $\omega$. The bases of this new frame are given by
\begin{subequations}
\begin{align}
\label{eX}
\bm{e}_X&=\big[\cos\omega,-\sin\omega,0\big],\\
\label{eY}
\bm{e}_Y&=\left[\cos\iota\sin\omega,\cos\iota\cos\omega,-\sin\iota\right],\\
\label{eZ}
\bm{e}_Z&= \left[\sin\iota\sin\omega,\sin\iota\cos\omega,\cos\iota\right] ,
\end{align}
\end{subequations}
in terms of the bases of the non-rotating frame, and $h_{+}$ and $h_{\times}$ can be produced in terms of $\bm{e}_X$ and $\bm{e}_Y$ building the transverse subspace as follows (\citealp{poisson2014gravity})

\begin{subequations}
\begin{align}
\label{h_plus1}
& h_+=\frac{1}{2}\big(e^j_X e^k_X-e^j_Y e^k_Y\big)h_{jk},\\
\label{h_cross1}
& h_{\times}=\frac{1}{2}\big(e^j_X e^k_Y+e^j_Y e^k_X\big)h_{jk}.
\end{align}
\end{subequations}

Now, we have enough material to construct $h_{+}$ and $h_{\times}$. Inserting Eqs. 
\eqref{IFxx}-\eqref{eZ} and \eqref{Qxxzz}-\eqref{Qyzzz} within definition \eqref{h_plus1}, after some simplifications, we find the plus polarization as 
\begin{align}
\nonumber
& h_+=\frac{1}{2} h_{0\text{M}}\left(1+\cos^2\iota\right)\cos 2\left(\Omega \tau +\omega\right)\\
\nonumber
&~~~~~- h_{0\text{F}}\sin\theta_0\Big[\sin\theta_0\left(1+\cos^2\iota\right)\cos 2\left(\Omega \tau+\omega+\varphi_0\right)\\
\label{h+}
&~~~~~~~~~~~~~~~~~~~~~~~~~~~~~~+\cos\theta_0\cos\iota\sin\iota\sin \left(\Omega \tau+ \omega+\varphi_0\right)\Big],
\end{align}
where
\begin{subequations}
\begin{align}
\label{h0M}
& h_{0\text{M}}=\frac{4 G\epsilon_{\text{M}} I_3\Omega^2}{c^4 R_{\text{d}}},\\
\label{h0F}
& h_{0\text{F}}= \frac{32 G A_0 \Omega^2}{ 21 c^4 R_{\text{d}}},
\end{align}
\end{subequations}
are the matter and B-field parts of the GW amplitude, respectively. 
In fact, $h_{0\text{F}}$ is the 1\tiny PN \normalsize correction to the well-known GWs amplitude of the deformed rotating star, $h_{0\text{M}}$, by virtue of the presence of the magnetic field. 
Definition $h_{0\text{F}}$ is one of our main results.
We also introduce two scale-free polarizations for this component as 
\begin{subequations}
\begin{align}
\label{H+M}
& H_{+\text{M}}= \frac{1}{2}\left(1+\cos^2\iota\right)\cos 2\left(\Omega \tau +\omega\right),\\
\nonumber
& H_{+\text{F}}= -\sin\theta_0\Big[\sin\theta_0\left(1+\cos^2\iota\right)\cos 2\left(\Omega \tau+\omega+\varphi_0\right)\\\label{H+F}
&~~~~~~~~~~~~~~~~~~~~~~~~+\cos\theta_0\cos\iota\sin\iota\sin \left(\Omega \tau+ \omega+\varphi_0\right)\Big].
\end{align}
\end{subequations}
$H_{+\text{M}}$ and $H_{+\text{F}}$ reveal the matter and B-field portions of the GW propagation in terms of the retarded time, respectively. To extract the cross polarization, we use the same method as explained before. So, for this component, we finally arrive at
\begin{align}
\nonumber
& h_{\times}= h_{0\text{M}}\cos\iota\sin 2\left(\Omega \tau+\omega\right)\\\nonumber
&~~~~~+ h_{0\text{F}}\sin\theta_0\Big[\cos\theta_0\sin\iota\cos\left(\Omega \tau +\omega+\varphi_0\right)\\\label{hcross}
&~~~~~~~~~~~~~~~~~~~~~~~~~~~~~~-2 \sin\theta_0\cos\iota\sin 2\left(\Omega \tau+\omega+\varphi_0\right)\Big].
\end{align}
Here, we also introduce two scale-free polarizations $H_{\times\text{M}}$ and $H_{\times\text{F}}$ as 
\begin{subequations}
\begin{align}
\label{HcrossM}
& H_{\times\text{M}}= \cos\iota\sin 2\left(\Omega \tau+\omega\right),\\
\nonumber
& H_{\times\text{F}}= \sin\theta_0\Big[\cos\theta_0\sin\iota\cos\left(\Omega \tau +\omega+\varphi_0\right)\\\label{HcrossF}
&~~~~~~~~~~~~~~~~~~~~~~~~-2 \sin\theta_0\cos\iota\sin 2\left(\Omega \tau+\omega+\varphi_0\right)\Big].
\end{align}
\end{subequations}

Therefore, according to Eqs. \eqref{h+} and \eqref{hcross}, the plus and cross polarizations of the transverse-tracefree part of the gravitational potential/the GW field, $h^{jk}_{\text{TT}}$, is obtained. In summary, by choosing a vacuum dipole model for magnetars, we find the direct contribution of the magnetic field to the GW field to the 1\tiny PN \normalsize order. In the following subsection, we study $H_{+\text{F}}$ and $H_{\times\text{F}}$ and compare these components with those of the matter part. We discuss the amplitudes of GW, i.e., $h_{0\text{M}}$ and $h_{0\text{F}}$, in Subsec. \ref{Application to magnetars}.

\begin{center}
\begin{figure}
\includegraphics[scale=0.64]{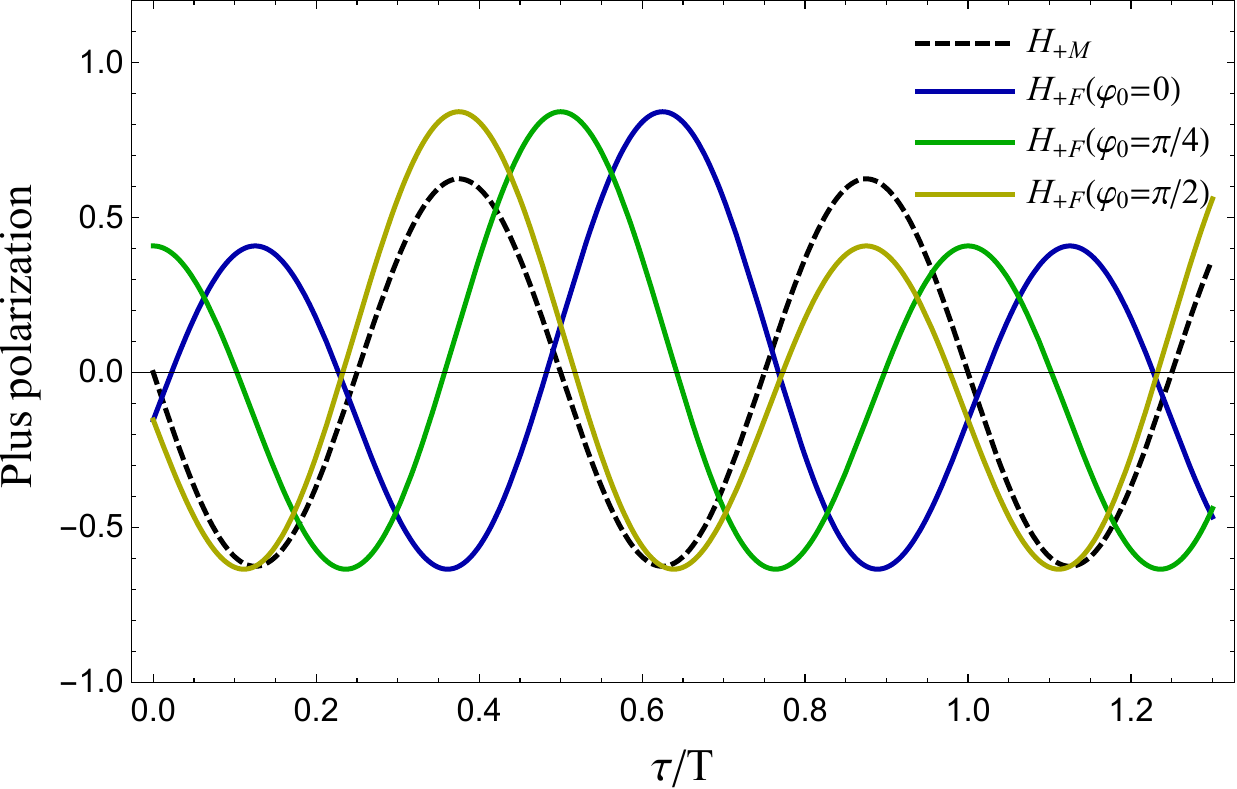}
\vspace*{.1cm}
\includegraphics[scale=0.64]{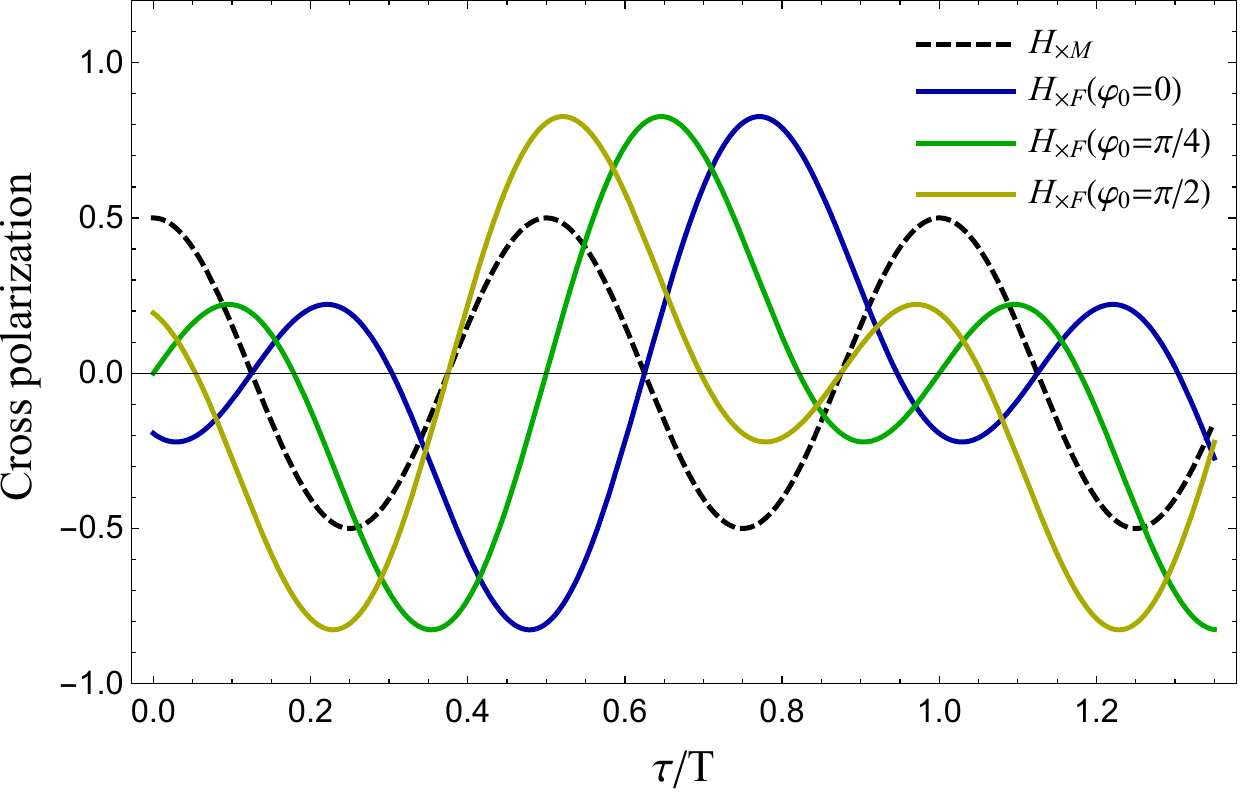}
\caption{Phase difference diagram. In the left and right panels, the colored curves exhibit $H_{+\text{F}}$ and $H_{\times\text{F}}$ in terms of different values of $\varphi_0$, respectively. In these panels, the dashed curve shows the corresponding scale-free polarization of the matter part. Also, we assume that $\iota=\pi/3$, $\omega=\pi/4$, and $\theta_0=\pi/4$.}\label{Fig varphi}
\end{figure}
\end{center}

\subsection{Discussion}\label{Discussion}

In the previous subsection, we have derived the structure of the plus and cross polarizations which contain the matter and B-field portions, and we have shown the dependency of these polarizations on the angles $\iota$, $\omega$, $\theta_0$, and $\varphi_0$.
In this subsection, in order to grasp the behavior of GW emitted from the rotating magnetized star, we attempt to study the scale-free polarizations.
We postpone our main discussion on the amplitude of GW signals until Subsec. \ref{Application to magnetars}, where we study the known magnetars and find $h_{0\text{F}}$ for each system.

We first evaluate the imprint of the angle $\varphi_0$ in the B-field part of GW. At first glance, it seems that this phase shift $\varphi_0$ is of no physical importance. However, as we will show, wave contributions of the matter and magnetic parts interfere with each other, and $\varphi_0$ controls the final intensity of the outcome. To simplify the following calculations, let us rewrite the scale-free polarization $H_{+\text{F}}$ as the following complex form
\begin{align}
\nonumber
& H_{+\text{F}}=\alpha_1 \left(e^{2 i(\Omega\tau+\omega+\varphi_0)}+e^{-2 i(\Omega\tau+\omega+\varphi_0)}\right)\\\nonumber
&~~~~~~~~~~~~~~~~~~~~~~~~~~~~~~~~~~~~~~~~~+i\alpha_2\left(e^{- i(\Omega\tau+\omega+\varphi_0)}-e^{i(\Omega\tau+\omega+\varphi_0)}\right)\\
\nonumber
&=e^{2 i(\Omega \tau+\omega)}\left(\alpha_1e^{2i\varphi_0}-i\alpha_2e^{-i(\Omega\tau+\omega-\varphi_0)}\right)\\\label{complex form}
&~~~~~~~~~~~~~~~~~~~~+e^{-2 i(\Omega \tau+\omega)}\left(\alpha_1 e^{-2i\varphi_0}+i\alpha_2e^{i(\Omega\tau+\omega-\varphi_0)}\right),
\end{align}
in which $\alpha_1=-1/2 \sin^2\theta_0 \big(1+\cos^2\iota\big)$ and $\alpha_2=-1/8 \sin 2\theta_0 \sin 2\iota$. By utilizing this form of the wave, we easily obtain the amplitude $\alpha$ and phase shift $f$ of $H_{+\text{F}}$ as
\begin{align}
\nonumber
\alpha&=\Big\arrowvert\alpha_1 e^{2i\varphi_0}-i \alpha_2 e^{-i(\Omega\tau+\omega-\varphi_0)}\Big\arrowvert\\
&=\sqrt{\alpha_1^2+\alpha_2^2-2\alpha_1\alpha_2\sin(\Omega\tau+\omega+\varphi_0)},
\end{align}
and
\begin{align}
f=\tan^{-1}\left(\frac{\alpha_1\sin 2\varphi_0-\alpha_2\cos (\Omega\tau+\omega-\varphi_0)}{\alpha_1\cos 2\varphi_0+\alpha_2\sin(\Omega\tau+\omega-\varphi_0)}\right),
\end{align}
respectively. Therefore, Eq. \eqref{complex form} can be rewritten as $H_{+\text{F}}=\pm\alpha \left(e^{i\left[2(\Omega\tau+\omega)+f\right]}+e^{-i\left[2(\Omega\tau+\omega)+f\right]}\right)$. Comparing this equation with the complex form of $H_{+\text{M}}$, i.e., $H_{+\text{M}}=1/4\left(1+\cos^2\iota\right)\left(e^{2 i(\Omega\tau+\omega)}+e^{-2 i(\Omega\tau+\omega)}\right)$ reveals that there is a time-dependent phase difference $f$ between these two parts of GWs. It should be mentioned that the phase shift $f$ is also a function of the angles $\iota$, $\omega$, $\theta_0$, and especially $\varphi_0$. Similarly, it can be shown that there is a phase difference 
\begin{align}
g=-\tan^{-1}\left(\frac{\alpha_3\sin(\Omega\tau+\omega-\varphi_0)+\alpha_4\cos 2\varphi_0}{\alpha_3\cos(\Omega\tau+\omega-\varphi_0)+\alpha_4\sin 2\varphi_0}\right),
\end{align}
between the matter and B-field parts of the cross polarization, i.e., $H_{\times\text{M}}$ and $H_{\times\text{F}}$ respectively. Here, $\alpha_3=1/4\sin 2\theta_0\sin\iota$ and $\alpha_4=-\sin^2\theta_0\cos\iota$. In the following, to study the pure effect of $\varphi_0$ on these phase shifts, we will fix the amount of $\iota$, $\omega$, $\theta_0$ and investigate $f(\tau,\varphi_0)$ and $g(\tau,\varphi_0)$.

We plot the plus and cross scale-free polarizations in the left and right panels of Fig. \ref{Fig varphi}, respectively. Hereafter, for simplification, we exhibit these components in terms of the dimensionless retarded time, $\tau/T$, which is divided to the rotational period of the star, $T=2\pi/\Omega$.
In Fig. \ref{Fig varphi}, the dashed curves belong to the matter part of GWs and the solid ones illustrate the B-field part. Comparing the dashed curve with the colored ones in both panels reveals that by changing $\varphi_0$, the matter and B-field parts of GWs can be partially constructive/destructive. So, for the various value of $\varphi_0$, partial interferences can happen.
It should be noted that, in this case, $\varphi_0$ is actually the angle between the magnetization and the long axis of the deformed star. In fact, as we mentioned earlier, NS is not necessarily deformed in the direction of the magnetic field lines and consequently $\varphi_0\neq 0$. So, under this physical circumstance, the partial interferences may happen.

\begin{figure*}
\centering
\includegraphics[scale=0.65]{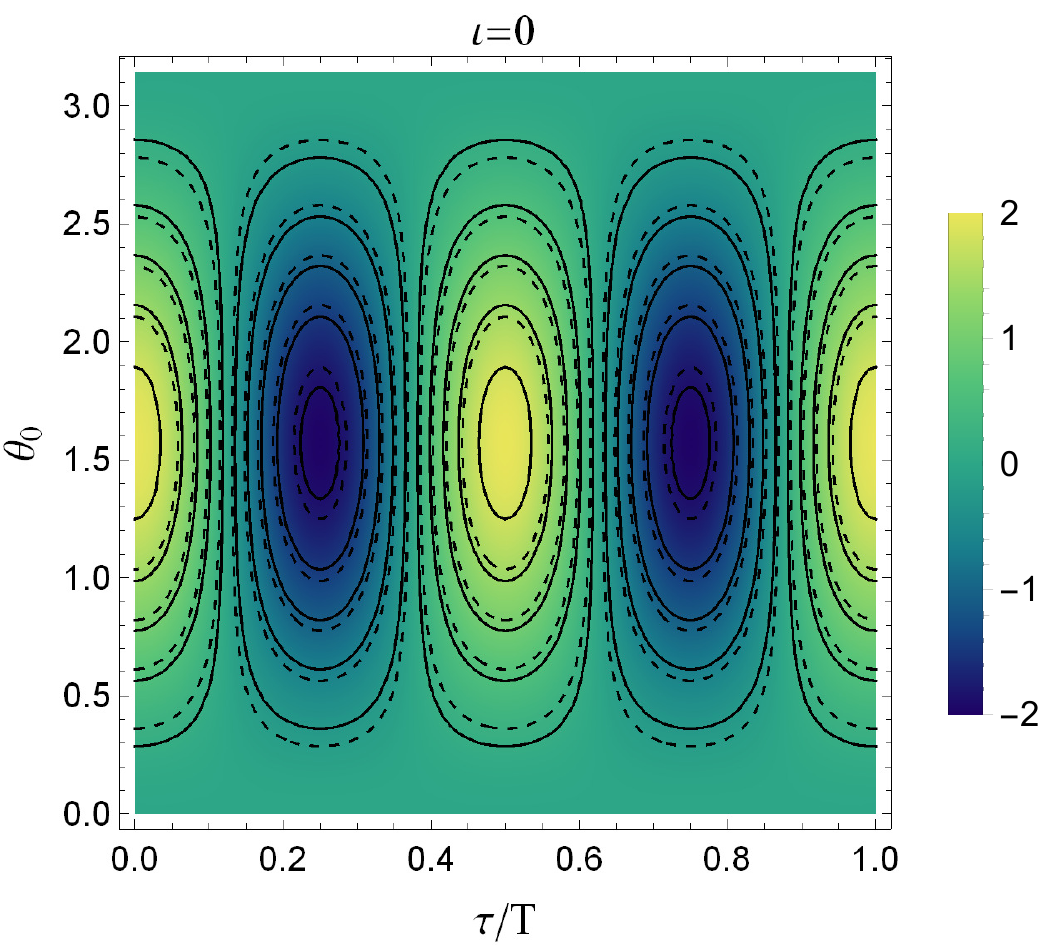}\hspace*{.15cm}
\includegraphics[scale=0.65]{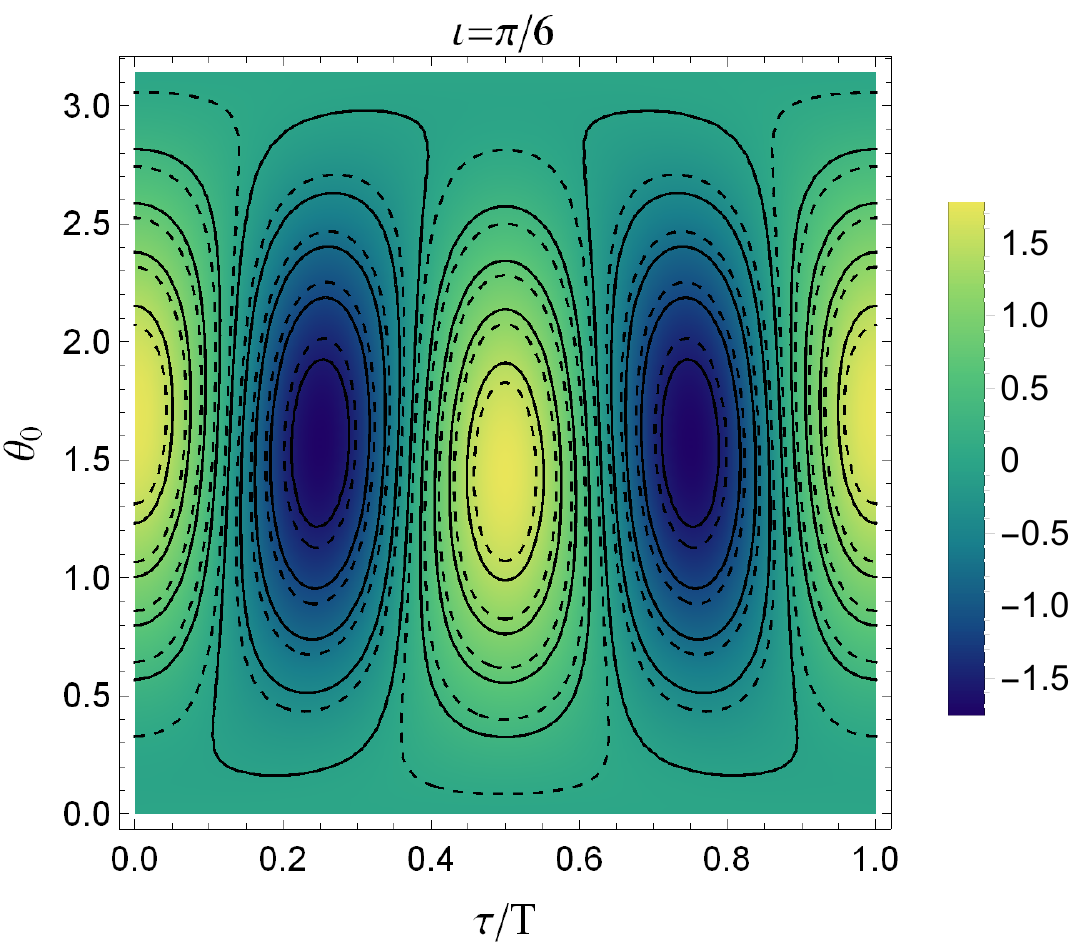}
\vspace*{.15cm}
\includegraphics[scale=0.65]{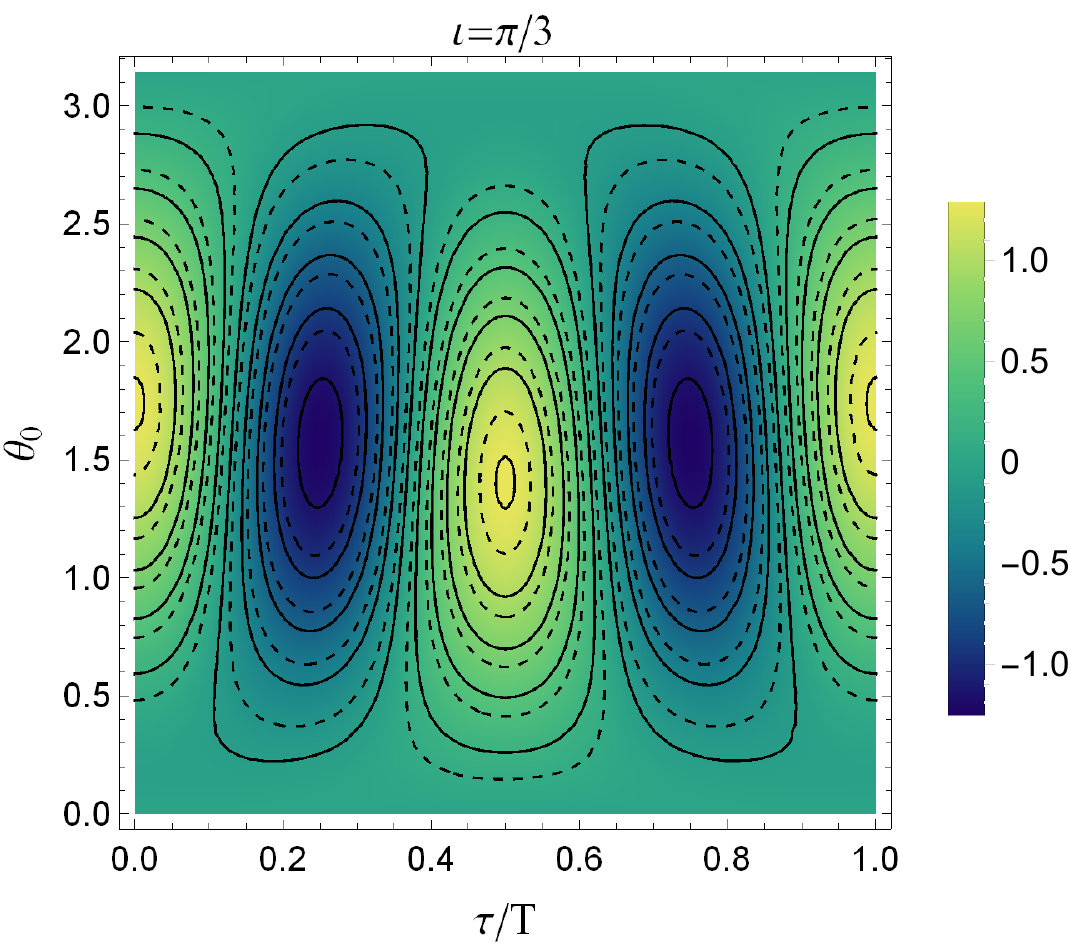}\hspace*{.15cm}
\includegraphics[scale=0.65]{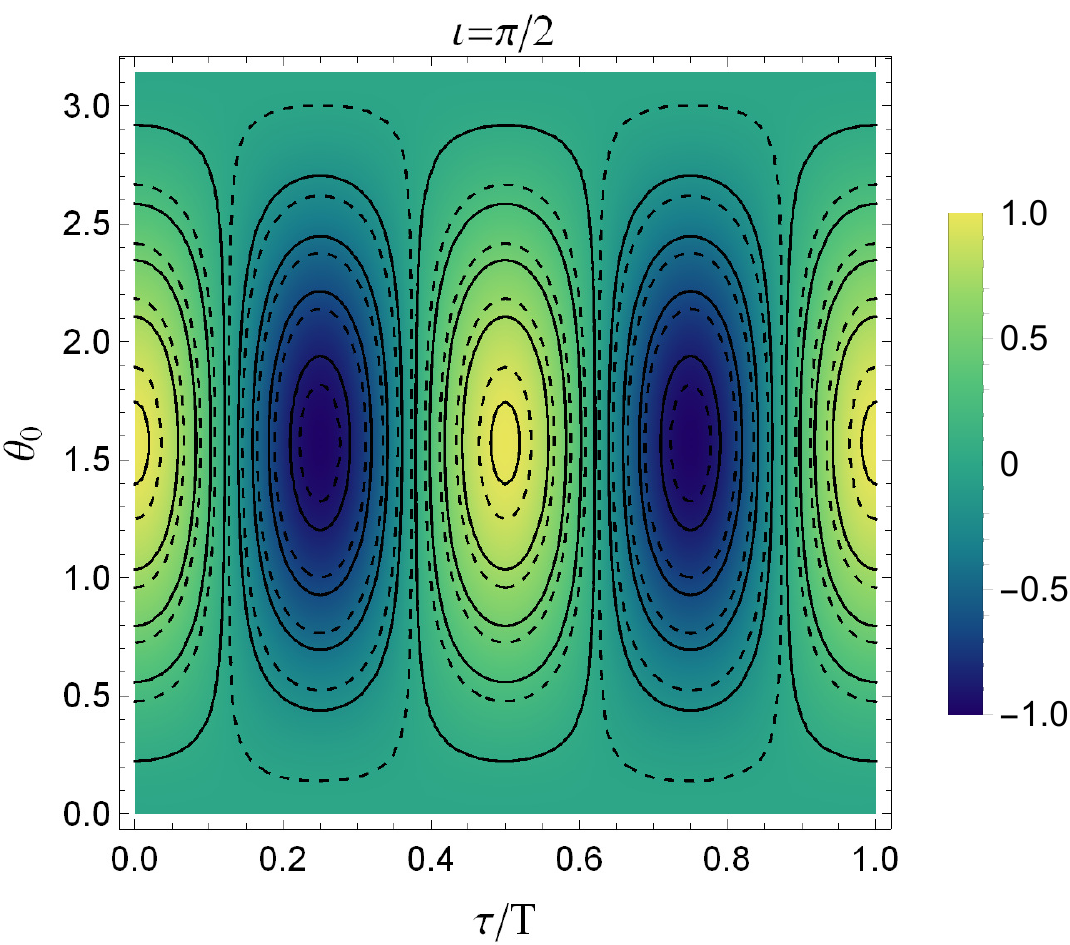}
\caption{Density plot of the B-field part of the scale-free plus polarization, $H_{+\text{F}}$. The colored bar displays the magnitude of this component of GWs. The dashed and solid curves represent the same-domain lines, i.e., $H_{+\text{F}}(\theta_0,\tau)=\textit{const}$. We assume that the total phase is $\omega+\varphi_0=\pi/2$. Each panel belongs to a different value for $\iota$. Here, $\theta_0$ is expressed in radians.}\label{Dens_plots-plus}
\end{figure*}
\begin{figure*}
\centering
\includegraphics[scale=0.65]{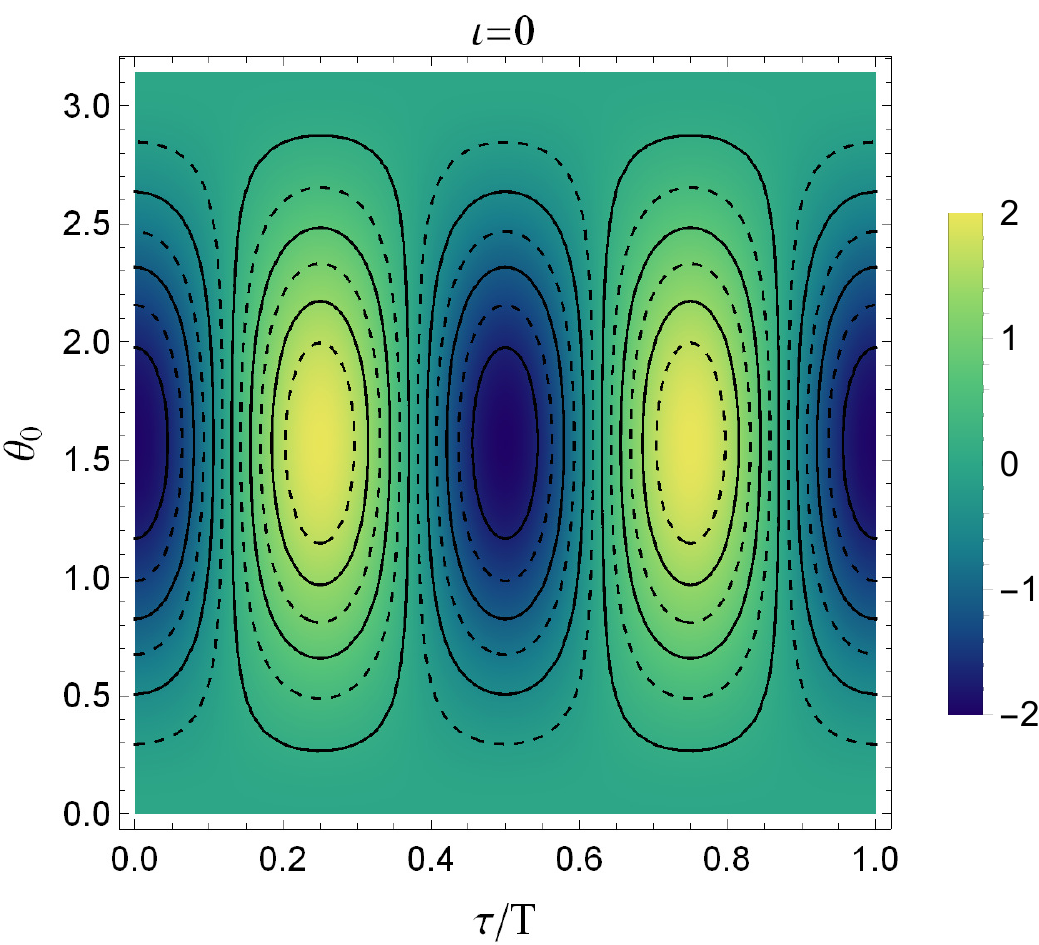}\hspace*{.15cm}
\includegraphics[scale=0.65]{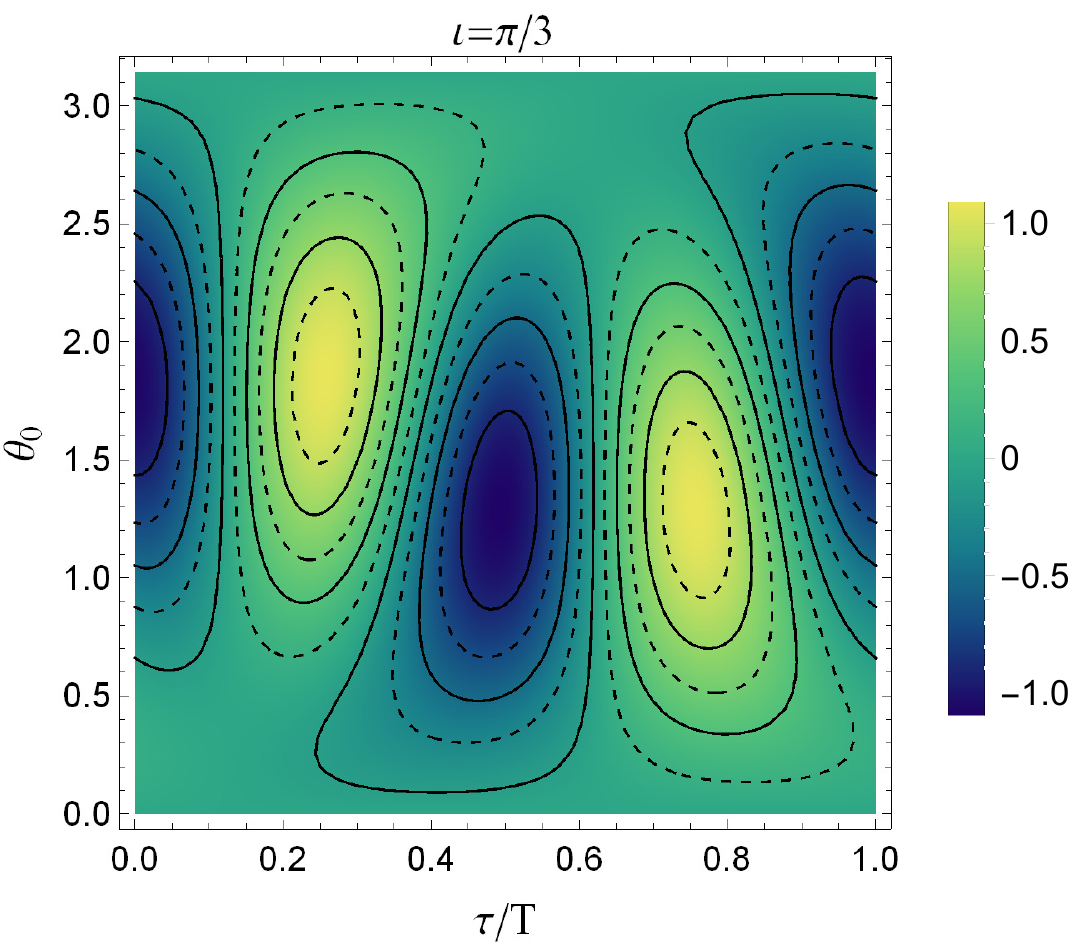}
\vspace*{.15cm}
\includegraphics[scale=0.65]{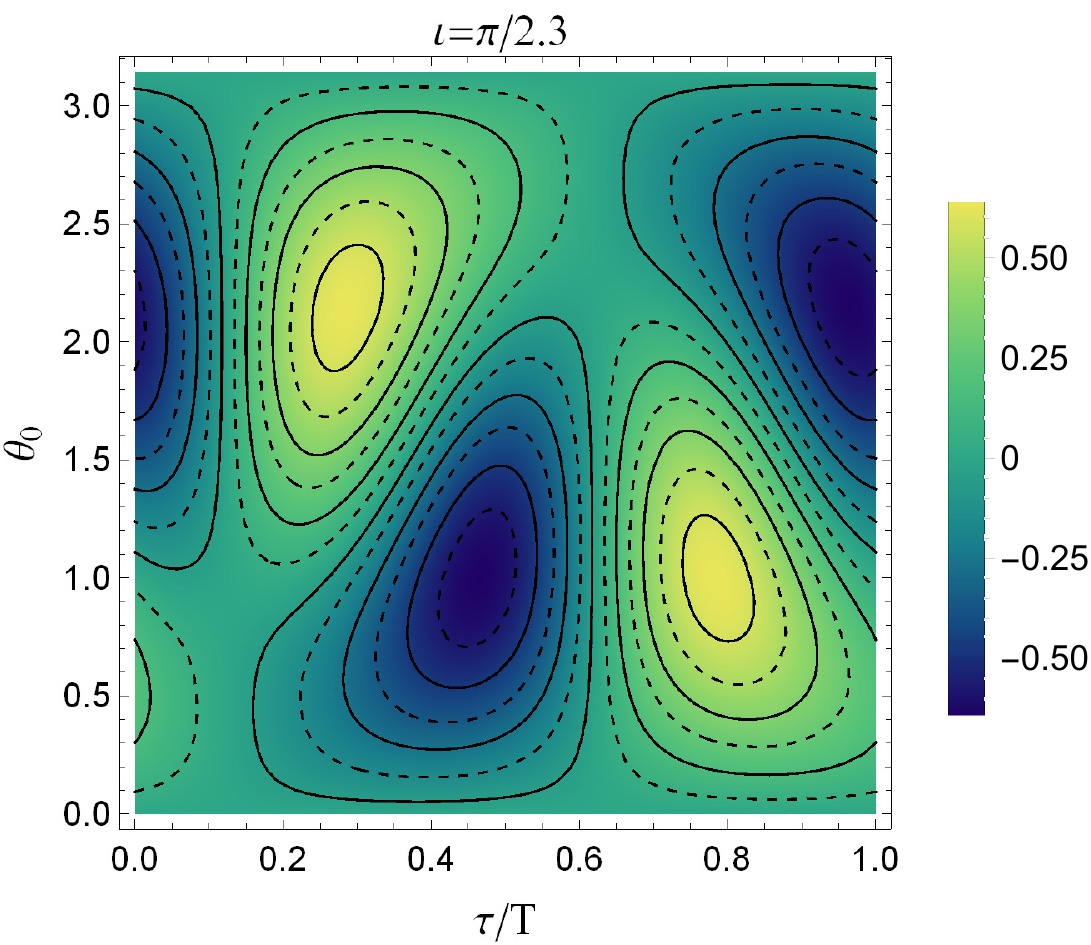}\hspace*{.15cm}
\includegraphics[scale=0.65]{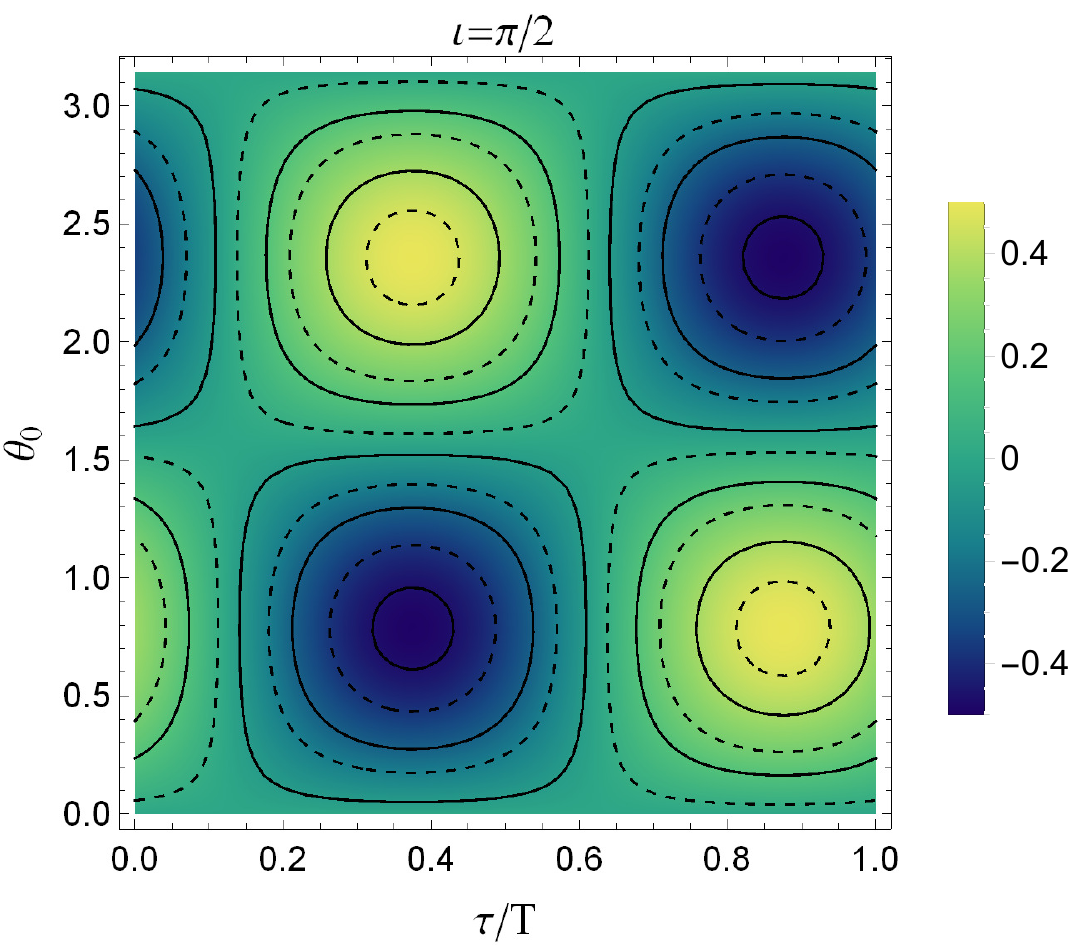}
\caption{Density plot of the B-field part of the scale-free cross polarization, $H_{\times\text{F}}$. In each panel, we have used a different value of $\iota$. The colored bar displays the magnitude of $H_{\times\text{F}}$. We also assume that the total phase is $\omega+\varphi_0=\pi/4$. Here, $\theta_0$ is expressed in radians.}\label{Dens_plots-cross}
\end{figure*}
\begin{center}
\begin{figure}
\includegraphics[scale=0.7]{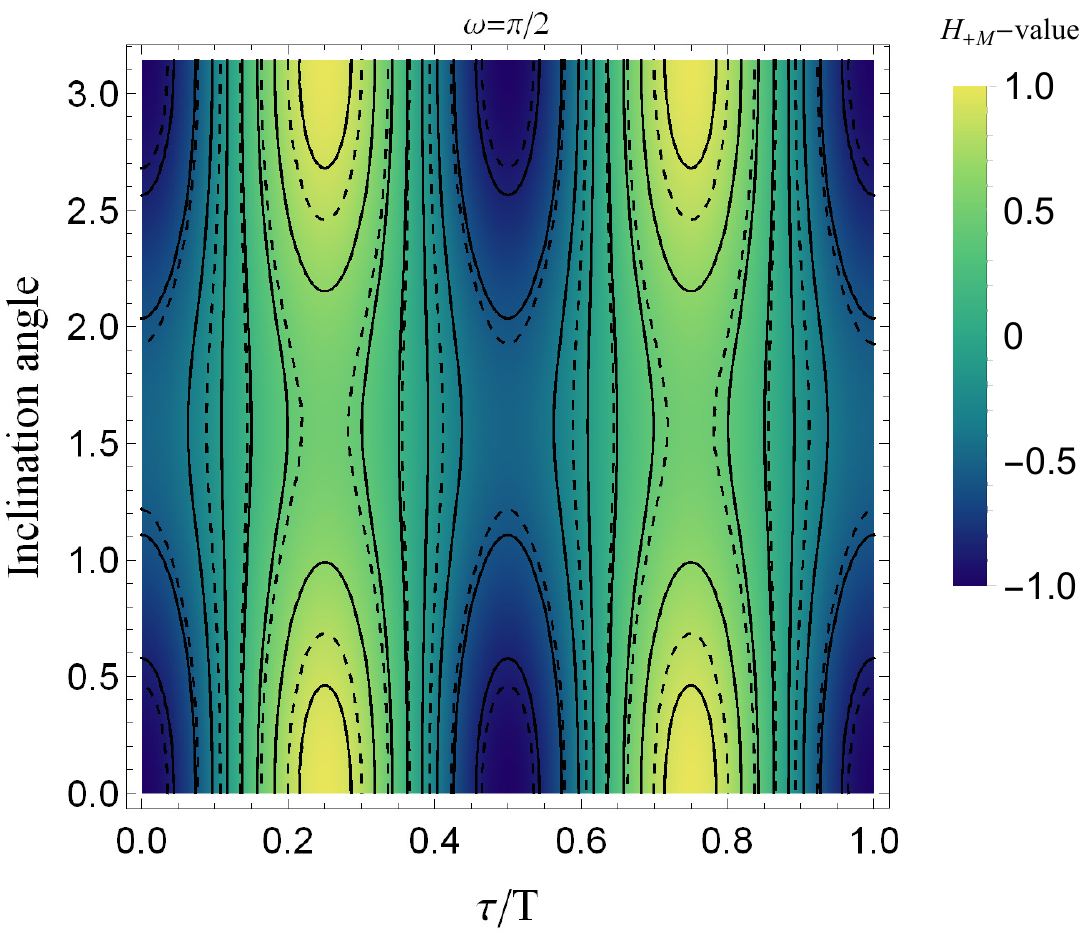}
\vspace*{.2cm}
\includegraphics[scale=0.7]{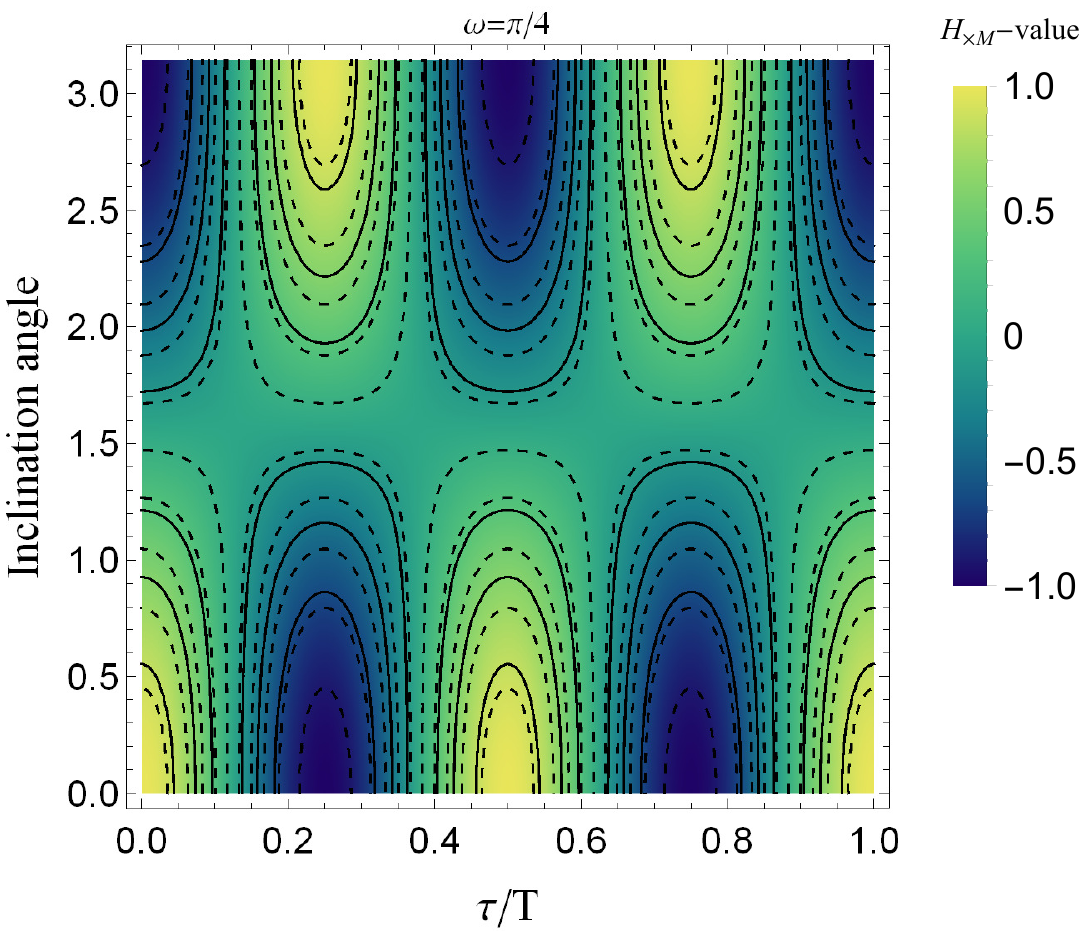}
\caption{Density plot of the matter part of the scale-free polarizations. The left and right panels exhibit $H_{+\text{M}}$ and $H_{\times\text{M}}$ in terms of the dimensionless retarded time and the inclination angle, respectively. Here, $\iota$ is expressed in radians. We recall that, here, the deformation is caused by other physical circumstances and it does not show the magnetically deformed star. \label{Dens_plots-M}}
\end{figure}
\end{center}
Another important parameter is the angle between the magnetic dipole $\bm{\mu}$ and the rotation axis of the star, i.e., $\theta_0$. It is clear that if $\theta_0=0$, then the magnetic field will not contribute to GW. To see the impact of this parameter, in Figs. \ref{Dens_plots-plus} and \ref{Dens_plots-cross}, we exhibit Eqs. \eqref{H+F} and \eqref{HcrossF} in terms of dimensionless retarded time and $\theta_0$, respectively. Also, in each figure, the evolution of the gravitational waveform is represented in terms of the inclination angle $\iota$.
In both figures, we can see that by increasing $\iota$ from $0$ to $\pi/2$, the contribution of the magnetic field to the amplitude of GW is reduced continuously. Moreover, as expected, this is also the case for the matter part of GWs. To see this effect, we also plot $H_{+\text{M}}$ and $H_{\times\text{M}}$ in terms of $\iota$ in Fig. \ref{Dens_plots-M}. The decline of the amplitude is more violent for the matter and B-field pieces of the cross polarization (See Fig. \ref{Dens_plots-cross} and the bottom panel of Fig. \ref{Dens_plots-M}). In fact, when the rotation axis is perpendicular to the line-of-sight, the GW detector receives very weak waves from both matter and B-field contributions.

As seen from Fig. \ref{Dens_plots-plus}, for $\iota=n\, \pi/2$ in which $n=0,1,2,\dots$, in the interval of time $\tau\in (0,T)$, i.e., in each rotation of the star, two GWs are propagated by virtue of the magnetic field of the star. This fact can be easily inferred from Eq. \eqref{H+F}. Apart from these angles, we observe one pattern of the wave for the plus polarization at each period (see the top right and  bottom left panels in Fig. \ref{Dens_plots-plus})
while for $H_{+\text{M}}$, for each value of $\iota$, two waves are emitted per each rotation of the star.
Furthermore, in Fig. \ref{Dens_plots-plus}, it is seen as long as the magnetic axis is parallel to the rotation axis, i.e., $\theta_0=n\pi$ where $n=0,1,2,\dots$, the B-field part of the plus polarization vanishes. Moreover, for all values of $\iota$, the strongest waves occur at the vicinity of $\theta_0=\pi/2$.

In Fig. \ref{Dens_plots-cross}, it is shown that for $\iota=n\pi$ in which $n=0,1,2,\dots$, the period of $H_{\times\text{F}}$ is half the rotational period of the star. So, as seen from the top left panel in this figure, two gravitational radiations are generated per each rotation of the star by virtue of the presence of the magnetic field associated with the star. Apart from these angles, each rotation of the star creates only one repeating pattern. However, for the corresponding component of the matter source, except at the vicinity of $\iota=\pi/2$, two waves are emitted at the time interval $0$ to $T$. This fact can be easily followed in Fig. \ref{Dens_plots-cross} and the bottom panel of Fig. \ref{Dens_plots-M}. 
It should be mentioned that, unlike the plus polarization, the maximum and minimum of the cross polarization are not necessarily created at the particular angle $\theta_0=\pi/2$.
As it is evident from Fig. \ref{Dens_plots-cross}, by increasing $\iota$, the peaks of the wave are displaced, and the amplitude of the cross polarization decreases continuously. Eventually, when $\iota=\pi/2$, it completely wipes out at $\theta_0=\pi/2$ (see the bottom right panel in Fig. \ref{Dens_plots-cross}). In this specific inclination angle, two patterns whose peaks are in $\theta_0\simeq \pi/4$ and $\theta_0\simeq3\pi/4$ are created with a  phase difference $\Delta=\pi/2$. 
In this case, as the plus component, at $\theta_0=0$ and $\theta_0=\pi$, no wave is generated by the magnetic field of the star.

As the last and the main point in this subsection, it should be noted that, at the inclination angle $\iota=\pi/2$, when the angle between the rotation and magnetic axes is about $\theta_0=\pi/4$ or $\theta_0=3\pi/4$, there is a possibility to receive a gravitational radiation from the magnetic star with a maximum magnitude $\arrowvert H_{\times\text{F}}\arrowvert\simeq 0.5$ for the B-field part of cross polarization. This fact can be easily grasped from the bottom right panel of Fig. \ref{Dens_plots-cross}. On the other hand, under this condition, the GW detectors cannot observe any wave generated by the corresponding matter contribution. See the bottom panel of Fig. \ref{Dens_plots-M}. Therefore, in this special situation, i.e., when $\iota=\pi/2$, every cross-polarization wave that is received from the rotating magnetar is only coming from its B-field part and matter part has no portion. This interesting fact helps us to discriminate between two different contributions of the matter and magnetic fields to GW signals. Also, this fact  moderately occurs for the plus polarization at $\theta_0=\pi/2$ whose magnitude is about $\arrowvert H_{+\text{F}}\arrowvert\simeq 1$.

\subsection{Realistic model}\label{Realistic model}

In this subsection, we attempt to describe GWs emitted from a more realistic rotating magnetar by applying the well-known force-free model already started by \cite{goldreich1969pulsar} (hereafter GJ). In this model, the region surrounding a rotating magnetic NS is studied. It is shown that these systems are not floating in a vacuum, and they have a dense magnetosphere rotating with the same frequency as the spinning star. In this model, the space around the star is divided into three zones and the magnetic field configuration is studied in each region.
These regions include the near zone\footnote{In order to distinguish between the near zone that we first encounter in the PN calculation and the case introduced in this division, hereafter, we call this region the light cylinder, and we keep the near zone concept in the PN context.}, which is restricted to the light cylinder with the radius $r\sin\theta=c/\Omega$ and the height $z=\pm c/\Omega$, the wind zone, the region between the light cylinder and a sphere with a radius $r\sim D/10$ where $D$ is the radius of the supernova shell, and the boundary zone, from the radius $D/10$ to $D$ beyond which there is the interstellar material.
It is also argued that two different currents can generate the magnetic field around the rotating magnetic NS. These currents are related to those inside the star and those within the corotating magnetosphere. The former is the main source of the poloidal component of the magnetic field, and the latter produces the toroidal field. Furthermore, it is shown that in the light cylinder, the major component is the poloidal one  determined approximately by the dipole magnetic field. On the other hand, in the wave zone the toroidal component dominates; and it is given by equation (14) in GJ.

Here, we utilize this model to specify the magnetic field configuration around the rotating magnetar in the near zone, and then describe GWs which can be emitted from this real system. To do so, let us first recall the size of the near zone, i.e., $\mathcal{R}$, in PN theory. As we see before, it is a sphere with a radius of order $\lambda_{\text{c}}=ct_{\text{c}}$ where $t_{\text{c}}$ is the characteristic time scale of the source. Here, for the current system, we consider that $t_{\text{c}}$ is equal to the rotational period of the magnetar, $T=2\pi/\Omega$, and consequently we have $\mathcal{R}=cT$. Moreover, as mentioned previously, the size of the light cylinder is of order $\sim c/\Omega$. Without loss of generality, we assume that this light cylinder is limited to a three-dimensional light sphere with a radius $b\simeq c/\Omega=c T/2\pi$. In fact, we introduce the cutoff parameter $b$ to determine the boundary of the corotating charge cloud. Therefore, by comparing $\mathcal{R}$ and $b$, one can deduce that this light sphere containing the corotating magnetosphere is completely embedded within the near zone. This directly means that the PN approximation works in a proper way for this system.
 
Now, we are in a position to determine the components of the magnetic field in the near zone. In GJ, the magnetic field of the rotating NS is considered to be symmetric around the rotational axis, i.e., $\theta_0\longrightarrow 0$ and $\bm{\mu}_{\text{m}}=\mu_{\text{m}}\bm{e}_z$ in our previous calculations. Although this symmetric configuration cannot radiate GW, we try to obtain the field piece of the mass quadrupole-moment tensor, $\mathcal{I}^{jk}_{\text{F}}$, for this system and estimate the magnitude of its components.
By applying this model, one can obtain
\begin{subequations}
\begin{align}
\label{Bi}
& B_{\text{in}}=\frac{2}{3}\mu_0\mu_{\text{m}}~~~~~~~~~~~~~~~~~~~~~~~~~~~~~~~~~~~~~~~~~~~~\text{for}~~r<a,\\
\label{Bp}
& B_{\text{p}}=\frac{\mu_0}{3}\frac{\mu_{\text{m}} a^3}{r^3}\left(3 \cos^2\theta+1\right)^{1/2}~~~~~~~~~~~~\text{for}~~a<r<b,\\
\label{Bt}
& B_{\text{t}}=-B_{\text{p}}\Big(\big(\frac{r}{b}\big)^2\sin^2\theta-1\Big)^{1/2}~~~~~~~~~~~~~~~~~\text{for}~~r>b,
\end{align}
\end{subequations}
for the dominant components of the magnetic field in each region in the near zone. As illustrated in GJ, the poloidal term, $ B_{\text{p}}$, is the major component in the corotating magnetosphere, i.e., $a<r<b$, This component, as shown in Eq. \eqref{Bp}, has a dipole configuration. In the wind zone, i.e., $r>b$, the dominant term is the toroidal component, $B_{\text{t}}$\footnote{For more detail, see section \textit{III} and equation (14) of GJ.}. According to this division of the near zone, we rewrite the integral \eqref{IjkF} as 
\begin{align}
\nonumber
\mathcal{I}^{jk}_\text{F}=&\frac{1}{2\mu_0c^2}\bigg[\int_{r<a} B^{2}_{\text{in}}\,x^jx^kd^3x\\
&+\int_{a<r<b} B^{2}_{\text{p}}\,x^jx^kd^3x+\int_{r>b} B^{2}_{\text{t}}\,x^jx^kd^3x\bigg].
\end{align}
In the following, we obtain the non-vanishing components of the above tensor. E.g., for the $x-x$ component of this tensor, we have
\begin{align}
\nonumber
&\mathcal{I}^{xx}_{\text{F}}
= \frac{1}{2\mu_0 c^2}\bigg[\int f_{\text{in}}\left(\theta,\varphi\right) d\Omega\int_0^a  r^4  d r +\int f_{\text{p}}\left(\theta,\varphi\right) d\Omega\times\\
\label{IXX}
&\Big(\int_a^b r^{-2} d r-\int_b^{\mathcal{R}}r^{-2} dr\Big)+\int f_{\text{t}}\left(\theta,\varphi\right)d\Omega\int_b^{\mathcal{R}} d r \bigg],
\end{align}
in which, as previous calculations, $f_{\text{in}}\left(\theta,\varphi\right)=B^{2}_{\text{in}}\sin^2\theta\cos^2\varphi$, and $d\Omega= \sin \theta d\theta d\varphi$ is an element of the solid angle. Also, here, $f_{\text{p}}\left(\theta,\varphi\right)=r^6 B^{2}_{\text{p}}\sin^2\theta\cos^2\varphi$, and $f_{\text{t}}\left(\theta,\varphi\right)=\left(r^6/b^2\right)B^{2}_{\text{p}}\sin^4\theta\cos^2\varphi$ are the $r$-independent poloidal and toroidal components, respectively. By inserting Eqs. \eqref{Bi}-\eqref{Bt} into Eq. \eqref{IXX}, one may straightforwardly integrate to obtain $\mathcal{I}^{xx}_{\text{F}}$ as follows
\begin{align}
\nonumber
\mathcal{I}^{xx}_{\text{F}}=8A_0\bigg(1-\frac{a}{21}\Big(\frac{38}{b}-\frac{14}{\mathcal{R}}-\frac{10\mathcal{R}}{b^2}\Big)\bigg),
\end{align}
where $A_0$ is introduced in Subsec. \ref{Simple model}. After deriving other non-vanishing components of this tensor and also, as mentioned before, freely omitting the $\mathcal{R}$-dependent terms, we finally arrive at
\begin{subequations}
\begin{align}
\label{IX}
&\mathcal{I}^{xx}_{\text{F}}= 8A_0\Big(1-\frac{38}{21}\frac{a}{b}\Big),\\
\label{IY}
&\mathcal{I}^{yy}_{\text{F}}= 8A_0\Big(1-\frac{38}{21}\frac{a}{b}\Big),\\
\label{IZ}
&\mathcal{I}^{zz}_{\text{F}}= 12A_0\Big(1-\frac{38}{21}\frac{a}{b}\Big).
\end{align}
\end{subequations}

As it is obvious from the above relations, in addition to $A_0$, these components also involve the ratio of the star radius $a$ to the cutoff parameter $b$, i.e., $a/b$. It should be noted that by a similar calculation, one can investigate that the ratio $a/b$ also appears in the components of $\mathcal{Q}_{\text{F}}^{jkab}$. To estimate the role of this term in the above equations, let us evaluate the magnitude of $a/b$. To do so, we choose a typical NS with a radius $a=12\, \text{km}$ and a rotational period $T=10\,\text{ms}$. For this realistic values, the ratio $a/b$ is of order $\sim 10^{-2}$. Even for this rapidly rotating NS, we see that this term is small.
Therefore, one can freely ignore the term including $a/b$ in comparison with the first term in Eqs. \eqref{IX}-\eqref{IZ}.

Considering the above arguments and recalling our discussion in Subsec. \ref{Simple model}, in which we show that the corresponding tensor of the simple model is only related to the term $A_0$, one can conclude that if this realistic system can radiate GW, the B-field part amplitude of the generated signal will be of the same order as the vacuum model in Eq. \eqref{h0F}. 
In fact, here, this realistic system can be well approximated by the simple dipole model. The validity of this approximation is also illustrated in \cite{gralla2017inclined}.
So, we can apply the previous results to describe the behavior of GW signals emitted by real magnetars. This means that our simple rotating NS model presented in Subsec. \ref{Simple model} can be effectively helpful to construct a viable model for a magnetar.

      \begin{table*}
         \caption{The GW amplitude of the magnetars}\label{table1}
       \begin{center}
      \centering\renewcommand{\arraystretch}{1.1}
      
         \begin{tabular}{lcccccc}
         
         \hline        
        
        Name  & $T$   & $\dot{T}$    & $B$     & $R_{\text{d}}$   & $h_{0\text{F}}$ \\
              & $\left(\text{s}\right)$& $\left(10^{-11}\text{s}\,\text{s}^{-1}\right)$& $\left(10^{10}\text{T}\right)$ & $\left(\text{kpc}\right)$  &  $\left(10^{-36}\right)$ &\\
         \hline
         \hline
         
        $\text{CXOU}\, \text{J}010043.1-721134$      & $8.02$      & $1.88$ &  $3.9$          & $62.4$      & $0.8$  \\
       
         $4\text{U}\,0142+61$              & $8.69$     & $0.20$ & $1.3$ & $3.6$         & $1.4$  \\

        $\text{SGR}\, 0418+5729$                    & $9.08$     & $4\times10^{-4}$  & $0.061$   & $\sim 2$         & $5.0\times10^{-3}$    \\
         $\text{SGR}\, 0501+4516$          & $5.76$  & $0.58$ &  $1.9$     & $\sim 2$     & $12.1$   \\
     
         $\text{SGR}\, 0526-66$ & $8.05$  & $3.8$   &  $5.6$   & $53.6$     & $2.0$  \\
        $1\text{E}\, 1048.1-5937$         & $6.46$  & $\sim 2.25$ & $3.9$& $9.0$  & $9.0$   \\
          $1\text{E}\,1547.0-5408$         & $2.07$  & $\sim 4.77$  &$3.2$&  $4.5$             & $117.9$                   \\
                $\text{PSR}\,\text{J}1622-4950$         & $4.33$  & $1.7$  &$2.7$&  $\sim 9$             & $9.6$                   \\
              $\text{SGR}\,1627-41$         & $2.59$  & $1.9$  &$2.2$&  $11.0$             & $14.6$                   \\
             $\text{CXOU}\,\text{J}164710.2-455216$         & $10.61$  & $< 0.04$  &$<0.66$&  $3.9$             & $0.2$    \\
              $1\text{RXS}\, \text{J}170849.0-400910$         & $11.00$  & $1.91$  &$4.6$&  $3.8$             & $10.2$    \\
           $\text{CXOU}\,\text{J}171405.7-381031$         & $3.83$  & $6.40$  &$5.0$&  $\sim 13.2$             & $28.6$    \\
           $\text{SGR}\,\text{J}1745-2900$         & $3.76$  & $0.66$  &$1.6$&  $\sim8.5$             & $4.7$    \\
           $\text{SGR}\,1806-20$         & $7.55$  & $\sim 49.5$     &$20$&  $8.7$             & $179.0$    \\
          $\text{XTE}\,\text{J}1810-197$         & $5.54$  & $0.78$     &$2.1$& $3.5$             & $9.1$    \\
           $\text{Swift}\,\text{J}1822.3-1606$         & $8.44$  & $0.031$     &$0.51$& $1.6$             & $0.5$    \\
           $\text{SGR}\,1833-0832$         & $7.57$  & $0.35$     &$1.6$& $\cdots$             & $9.9$    \\
           $\text{Swift}\,\text{J}1834.9-0846$         & $2.48$  & $0.80$     &$1.4$& $4.2$             & $16.8$    \\
           $1\text{E}\,1841-045$         & $11.78$  & $3.93$     &$6.9$& $8.5$             & $9.0$    \\
          $\text{SGR}\,1900+14$         & $5.20$  & $9.2$     &$7.0$& $12.5$             & $32.2$    \\
           $1\text{E}\,2259+586$         & $6.98$  & $0.05$     &$0.59$& $3.2$             & $0.5$    \\
                  \hline
        \end{tabular}
       \end{center}
       \vspace{1ex}
       
       {\raggedright Here, we assume that $a=10\,\text{km}$, $M=1.4\,M_{\odot}$, and $I=1.1\times 10^{38}\text{kg}\,\text{m}^2$. For each system, we have calculated the B-field part of the GW amplitude in the fifth column. It should be noted since the distance of $\text{SGR}\,1833-0832$ is not exactly specified in the literature, here as an estimation, we assume that $R_{\text{d}}$ is of the order $\sim 1\text{kpc}$.\par} 
       \end{table*}

\subsection{Application to magnetars}\label{Application to magnetars}
As we mentioned in the introduction, the effect of the magnetic field on GWs is studied in \cite{bonazzola1996gravitational} (hereafter BG). On the other hands, in BG, GWs are emitted from a rotating NS deformed by the presence of its own magnetic field. It is shown that this deformation is described by an ellipticity parameter given by equation (33) of BG
\begin{align}\label{epsilon}
\epsilon=\frac{5\pi}{4\mu_0 G}\frac{B^2 a^4}{ M^2},
\end{align}
where $M$ is the mass of the star. In fact, this formula reveals the contribution of the magnetic field to the deformation of NS, and consequently the emission of GW signals to the leading order.
It should be recalled that by considering the influence of the magnetic field on the matter distribution inside of NS, the equation \eqref{epsilon} is derived. So, we again call it the indirect effect of the magnetic field on the ellipticity. On the other hand, here, we examine another kind of ellipticity due to the presence of the inclined magnetic field which is not parallel to the rotation axis, not because of its effect on the dynamics of matter inside the star. From this perspective, as mentioned before, our analysis has a significant difference with the precedent works.

In the following, by using definitions \eqref{h0M} and \eqref{h0F}, we attempt to obtain magnetic ellipticity $\epsilon_{\text{F}}$ which is directly induced by the presence of the strong magnetic field inside and around the magnetar to the 1\tiny PN \normalsize order. In order to take into account the influence of the deviation of the magnetic and rotation axes on the magnetic ellipticity, here, we insert $\sin \theta_0$ term appearing in the B-field part of Eqs. \eqref{h+} and \eqref{hcross} into definition $h_{0\text{F}}$.
This effect on the amplitude of GWs is completely explained in Subsec. \ref{Discussion}.
Therefore, we rewrite Eq. \eqref{h0F} as  
\begin{align}
\label{h0new}
h_{0\text{F}}=\frac{32}{21}\color{black} \frac{ G A_0 \Omega^2}{ c^4 R_{\text{d}}}\sin\theta_0.
\end{align}
At this stage, comparing Eqs. \eqref{h0M} and \eqref{h0new}, we deduce that $\epsilon_{\text{F}}=8A_0\sin\theta_0/21 I_3$.
This parameter, as Eq. \eqref{epsilon}, can be also rewritten in terms of the magnetic field, radius, and mass of a star as
\begin{align}
\epsilon_{\text{F}}=\frac{4\pi }{189 c^2 }\color{black}\frac{\mu_0\mu_{\text{m}}^2 a^3}{M}\sin\theta_0=\frac{\pi}{21  c^2}\color{black}\frac{B^2 a^3}{\mu_0 M}\sin\theta_0,
\end{align}
after substituting the moment of inertia of the star with mass $M$, $I_3=\left(2/5\right) M a^2$, and the definition $A_0$.
In the above relation, we also use this fact that $B=\left(2/3\right)\mu_0\mu_{\text{m}}$.
 As expected, this parameter does not play any role in the Newtonian order, and its leading effect appears in the 1\tiny PN \normalsize order. Furthermore, when the magnetic field lines are symmetric around the rotation axis, i.e., $\theta=0$ or $\theta=\pi$, the ellipticity $\epsilon_{\text{F}}$ vanishes. And consequently, as discussed in Subsec. \ref{Discussion}, this symmetric magnetic field configuration does not generate any wave.
This issue is also reported in studies of the indirect effect of the magnetic field on GWs (\citealp{bonazzola1996gravitational}). One may take the ratio $\epsilon_{\text{F}}/\epsilon$ as a parameter representing the relative importance of indirect and direct effects of magnetic field on the generation of GWs. It is easy to show that $\epsilon_{\text{F}}/\epsilon\simeq 0.02 R_{\text{S}}/a$, where $R_{\text{S}}$ is the Schwarzschild radius of the magnetar. We know that $R_{\text{S}}<a$ for magnetars, and consequently one may infer that $\epsilon_{\text{F}}/\epsilon<0.02$. This means that the direct impact of the magnetic field on GW is small compared to its indirect effect. Therefore, the main question that arises is: does this weak signal associated with the direct impact of the magnetic field lie within the detection resolution of the current GW detectors? In what follows, we try to answer this question.

In order to numerically estimate the order of the magnitude of the ellipticity parameter, we choose a typical magnetar with $B=10^{15}\,\text{G}$, $M=1.4 M_{\odot}$, and $a=10\,\text{km}$. Then, we have 
\begin{align}
\epsilon_{\text{F}}\simeq 4.7\times 10^{-9}\color{black}\sin\theta_0\bigg(\frac{B}{10^{15}\,\text{G}}\bigg)^2\bigg(\frac{a}{10 \,\text{km}}\bigg)^3\bigg(\frac{1.4 M_{\odot}}{M}\bigg).
\end{align}
One can show that for this typical magnetar, the indirect magnetic ellipticity, i.e., Eq. \eqref{epsilon}, is about $\sim 6\times 10^{-7}$, confirming our above mentioned description that $\epsilon_{\text{F}}/\epsilon$ is always smaller than $0.02$. 

Now, we turn to assess the B-field part of the GW amplitude from magnetars. In the following, we first estimate $h_{0\text{F}}$ of the rapidly rotating magnetic NS known as the millisecond magnetar, e.g., see \cite{usov1992millisecond}, and we then study this amplitude which can be produced by the 21 well-known magnetars classified in \cite{olausen2014mcgill}. The results are summarized in Table. \ref{table1}.
By rewriting Eq. \eqref{h0F} in terms of $T$ and also using definition $A_0$, one can obtain that
\begin{align}
h_{0\text{F}}=\frac{32}{105}\color{black}\frac{\pi^3 G}{\mu_0 c^6 }\frac{B^2 a^5}{T^2 R_{\text{d}}}.
\end{align}
Here, to describe the millisecond magnetar, we consider the typical radius $12\,\text{km}$ and mass $1.4 M_{\odot}$, e.g., see \cite{dall2009early}.
Therefore, for this system with $B=10^{15}\,\text{G}$, $a=12\,\text{km}$, and $T=10\,\text{ms}$, we arrive at
\begin{align}
h_{0\text{F}}\simeq 5.5\times 10^{-28}\color{black}\bigg(\frac{B}{10^{15}\,\text{G}}\bigg)^2\bigg(\frac{a}{12 \,\text{km}}\bigg)^5\bigg(\frac{10 \,\text{ms}}{T}\bigg)^2\bigg(\frac{1\,\text{kpc}}{R_{\text{d}}}\bigg),
\end{align}
where we also assume this system can exist at $R_{\text{d}}=1\,\text{kpc}$. 
This is the order of the magnitude of the nearest known magnetar (\citealp{olausen2014mcgill}).
Furthermore, for this system with a moment of inertia of order $I_3=\left(2/5\right)M a^2\simeq 10^{38}\,\text{kg}\,\text{m}^2$ and a typical ellipticity parameter $\epsilon_{\text{M}}\simeq10^{-6}$, one can obtain that the matter part of the GW amplitude, $h_{0\text{M}}$, is of order $\sim 4.1\times10^{-26}$.
So, as deduced from these estimations, the B-field part of GWs released in the early stages of a magnetar's life can be as much as one-hundredth of the signals due to the deformed rotating NSs. 
Obviously, the amplitude $h_{0\text{F}}$ will increase for the models that predict a magnetic field of magnitude more than $10^{15}\,\text{G}$ inside the rapidly rotating magnetars, for example, see \cite{thompson1993neutron}.
Moreover, it is recently shown that the magnetic field is not, in general, a dipole and its multipole structure can more accurately describe a millisecond magnetar (\citealp{riley2019nicer,lockhart2019x}). 
In this case, it is also shown that the strength of the magnetic field on the surface of these systems can increase (\citealp{gralla2016pulsar}). In fact, the presence of the multipole structure can enhance the B-field contribution to GWs.
So, to find more accurate results, the dipole field assumption should be modified
in the future works. Regarding the complex nature of the PN approach, as the first step, we have restricted ourselves to the most simple model. Of course, this model is widely implemented in the literature, e.g., see (\citealp{turolla2015magnetars}).

As a final task in this subsection, let us gather the same calculation for the 21 well-known magnetars in Table. \ref{table1}. In this table, from left to right, the name, period, its derivative, magnitude of the magnetic field, and distance of the magnetars are expressed respectively, and the B-field part of the GW amplitude is displayed in the last column. As seen, the wave amplitude for most systems is extremely small and is of order $h_{0\text{F}}\sim 10^{-35}$. According to these results, the strongest amplitude is related to $\text{SGR}\,1806-20$. 
Therefore, according to the above discussion, it is seen that the magnetic field associated with these real systems detected until now produces very weak GW signals and its contribution in the final wave is insignificant. It should be mentioned that, here, the magnetic field inside these systems is predicted to be less than $10^{15}\,\text{G}$, while, some theories state that the strength of the magnetic field can be higher, and be of order $\sim 10^{18}\,\text{G}$ (\citealp{usov1992millisecond,nakamura1998model,kluzniak1998central,wheeler2000asymmetric}). So, it is possible if this system exists, its magnetic field can emit the "loud" signals which are detectable by the GW detectors.
On the other hand, in the formation scenario of magnetars, it is illustrated that they should be born with $\sim 1\text{ms}$ spin period. E.g., see \cite{dall2009early}. Therefore, if the magnetar candidates were born with a very fast rotating period, their B-field portion would have emitted a considerable GW signal. In the following,
we focus our attention on the millisecond magnetars and study the
amplitude and detectability of these systems.

\section{AMPLITUDE AND DETECTABILITY}\label{Detectors}

\begin{figure}
\begin{center}
\centering
\includegraphics[scale=0.63]{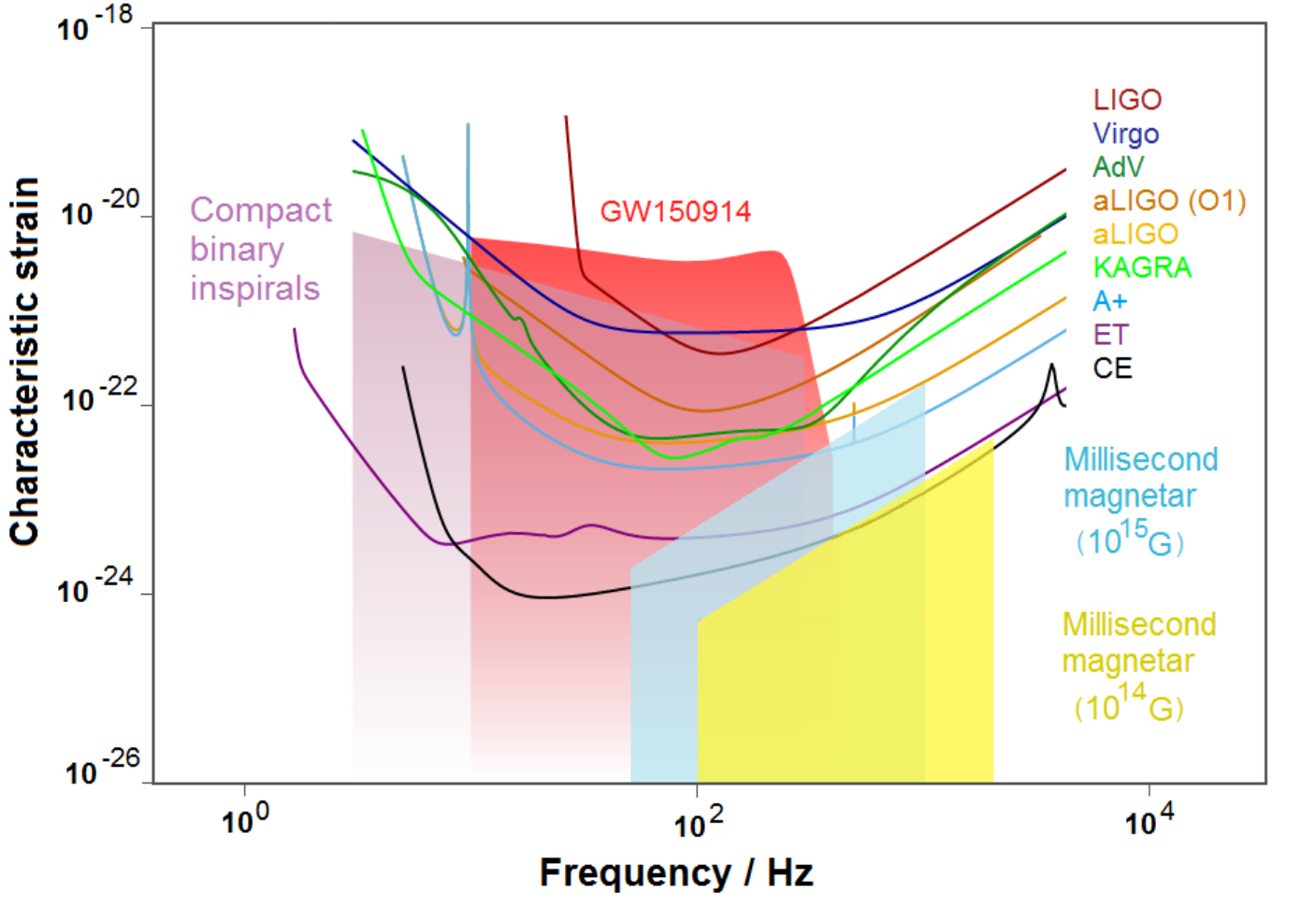}
\caption{The characteristic strain in terms of the frequency. The solid
colored lines represent the sensitivity curves of the GW detectors and the colored areas show the expected GW strain of the astrophysical systems. To compare our results with the promising and detected sources of GWs, we also illustrate the signal strain of the massive binaries, compact binary, GW150914, and so on. This plot is produced with a useful tool at \url{http://gwplotter.com} 
introduced in \protect\cite{moore2014gravitational}.}\label{characteristic strain}
\end{center}
\end{figure}

In this section, we investigate the characteristic strain of the GW emissions generated by the pure B-field part of the possible highly magnetized astrophysical sources.  Then, we compare it with the sensitivity range of the current and future GW detectors. To do so, let us rewrite the relation between the characteristic strain and the GW amplitude. See figure (A1) and equation (35) of \cite{moore2014gravitational}.  By considering the characteristic strain of the detectors $h_{\text{c}}$ in the frequency range $f$, we have 
\begin{align}\label{hc}
h_{\text{c}}\left(f\right)=\sqrt{\frac{2f^2}{\dot{f}}}h_0,
\end{align}
in which $\dot{f}$ shows the variation of the frequency in terms of the time. By using the above relation and considering $h_{\text{c}}$ of the detectors, one can conclude that after several years of integration, the LIGO and Virgo can observe the wave with the minimum amplitude of order $10^{-26}$ in the frequency range $10^2\,\text{Hz}$. In Fig. \ref{characteristic strain}, we attempt to display the characteristic strain of GW emitted only by the B-field part of the millisecond magnetar which is studied in the previous section. In this case, by considering $h_{\text{c}}$ only for the magnetic contribution in the frequency range $f$, Eq. \eqref{hc} reduces to 
\begin{align}\label{hcc}
h_{\text{c}}\left(f\right)=\sqrt{\frac{2f^2}{\dot{f}}}h_{0\text{F}}.
\end{align}
Here, we also illustrate the sensitivity curves of the GW detectors in order to indicate the signals which are strong enough to be observed by detectors. As it is obvious from Fig. \ref{characteristic strain}, due to the very low characteristic strain of the real magnetars which are introduced in Subsec. \ref{Application to magnetars}, they cannot be observed by the current detectors in the frequency band $10^{-1}\,\text{Hz}<f<1\,\text{Hz}$.

Before turn to obtain $h_{\text{c}}$ of the millisecond magnetars, let us briefly review two scenarios of forming a millisecond magnetar. It is stated that this system can form in the core-collapse supernovae (SNe) or can be the remnant of BNS mergers (\citealp{dall2018neutron}). Based on these two scenarios, we consider two distance scales, i.e., intra-galactic and extra-galactic distances, where a millisecond magnetar may exist.

We first obtain $h_{\text{c}}$ of a intra-galactic millisecond magnetar.
Here, we assume the typical millisecond magnetar with $B=10^{15}\,\text{G}$ and $R_{\text{d}}=1\,\text{kpc}$. We also assume that the time of integration is of the order of a few days. In fact, we consider that this time is equal to the spin-down timescale of the millisecond magnetars $t_{\text{sd}}$ given by $t_{\text{sd}}\simeq 4.7\,\text{day}\,B^{-2}_{14}T^2_{\text{ms}}$ in which $T_{\text{ms}}=T_0/1\,\text{ms}$ is the scaled birth spin period of NS $T_0$ and $B_{14}=B_{\text{d}}/10^{14}\,\text{G}$ is the surface dipole magnetic field $B_{\text{d}}$ that is normalized to $10^{14}\,\text{G}$ (\citealp{metzger2017millisecond}).
In this case, one may write $\dot{f}\simeq f/t_{\text{sd}}$, and consequently we have $h_{\text{c}}(f)\simeq 5.0\times10^{-27} f^{3/2}$.
We find that the characteristic strain of this intra-galactic millisecond magnetar is of the order $1.8\times10^{-24}\lesssim h_{\text{c}} \lesssim 1.6\times10^{-22}$ in the frequency band $50\,\text{Hz} \leqslant f \leqslant 10^3\,\text{Hz}$. The characteristic strain of this millisecond magnetars in terms of the frequency is displayed by a light blue area in Fig. \ref{characteristic strain}.
As shown, this area enters the sensitivity range of the several GW detectors, i.e., some part of this area lies above the sensitivity curve of the detectors. 
This means that the signal-to-noise ratio (S/N) 
can be large enough to be detected (\citealp{moore2014gravitational}). So, this millisecond magnetar which is in its early stages of evolution can emit loud signals that are audible to these detectors.
But as grasp from this light blue box, if the magnetar candidates were born with a lower frequency of order $10^{2}\,\text{Hz}$, S/N  would dramatically reduce. Furthermore, at the low-frequency band, $<10\,\text{Hz}$, one can show that only a super-strong magnetar with $B>10^{16}\,\text{G}$ may generate strong GWs. But these waves are beyond the frequency band of the detectors like CE and ET and they may be so hardly detected by DECIGO and BBO.

We should keep in mind that only millisecond magnetars with $B\simeq 10 ^{15}\,\text{G}$ can generate such loud signals. To clarify this issue, the characteristic strain of a similar millisecond magnetar with $B=10^{14}\,\text{G}$ is also exhibited in Fig. \ref{characteristic strain}. As it is obvious, considering the yellow area reveals that S/N is insignificant for this system.
Furthermore, in order to have a more realistic perspective, we should consider the millisecond magnetar birth rate. In \cite{kalogera2001coalescence}, it is estimated that the galactic BNS coalescence rate is of the order $\simeq 10^{-6}-5\times10^{-4}\,\text{yr}^{-1}$.
Moreover, the core-collapse SN rate is estimated of $2.30\pm 0.48$ SNe per century within the Milky Way Galaxy (\citealp{li2011nearby,timmes1997constraints}).
So, one can conclude that in the intra-galactic distances, the millisecond magnetar birth rate may lie in the range $(2.30\pm 0.48)\times10^{-2}\,\text{yr}^{-1}$.

Next, by considering the definition \eqref{hcc}, one can show that the characteristic strain of an extra-galactic millisecond magnetar which can be formed in the core-collapse of SNe, even for the super-strong magnetic field, is of order $<10^{-24}$. Therefore, the magnetic field of this system produces very faint GW signals so that they cannot be observed by the future third-generation detectors like ET and CE. In this scenario of forming a millisecond magnetar, its birth rate is of the order $\gtrsim 0.01\,\text{yr}^{-1}$ within the local extra-galactic distance $4\,\text{Mpc}$ (\citealp{dall2018neutron}). Moreover, it is shown that the event rate of BNS mergers is approximately of order $\sim \left(0.9-13\right)\times 10^{-4}\,\text{yr}^{-1}$ within $4\,\text{Mpc}$ (\citealp{dall2018neutron}). Therefore, millisecond magnetars are expected to form by rate $\gtrsim 0.01$ event per year within this distance.

\section{Summary and conclusion}\label{Summary and conclusion}

In this paper, we have attempted to study the direct effect of the magnetic field on GWs.
In Sec. \ref{General review}, we have studied the multipole moments constructing the PN gravitational potential and estimated the PN order of these moments. To do so, we have utilized the result of \cite{nazari2018post} which is briefly reviewed in Subsec. \ref{Post-Newtonian charged fluid}.
It has been shown that the direct role of the magnetic field appears, at least, in the 1\tiny PN \normalsize order. To determine the magnitude of this direct influence, we have computed the quadrupole-moment and 4-pole-moment tensors (i.e.,  $\mathcal{I}^{jk}_{\text{F}}$ and $\mathcal{Q}^{jkab}_{\text{F}}$, respectively) for NSs which have a strong magnetic field.
One can easily investigate that by dropping the B-field part of our derivations, the ordinary formula of the deformed rotating NS can be recovered. 
 
In Sec. \ref{rotating Magnetar}, the rotating magnetar has been studied. Because the configuration of the magnetic field is unknown in this system, we have chosen the simplest model, the vacuum dipole model, and obtained the B-field part of the quadrupole-moment and 4-pole-moment tensors. Moreover, we derived the plus and cross polarizations which contain the B-field part in Subsec. \ref{Simple model}. In fact, we have shown, for the first time, that the presence of the magnetic field can directly alter both the waveform, $H_{+,\times}$, and amplitude, $h_{0}$, of GW signals to 1\tiny PN \normalsize order. Then, in Subsec. \ref{Discussion}, we have studied the scale-free polarizations $H_{+\text{F}}$ and $H_{\times\text{F}}$ which govern the behavior of the PN waveforms emitted from the magnetic field of this system. To do so, we have evaluated the dependency of these polarizations on the different angles, e.g., the inclination angle $\iota$ and the angles between the magnetization and $z$ and $x$-axes, i.e., $\theta_0$ and $\varphi_0$, respectively. 

It turns out that for the various value of $\varphi_0$, the partial constructive and destructive interferences of two matter and B-field waves can occur.
We have illustrated when $\iota=0$ ($\iota=\pi/2$), the GW detectors receive very strong (weak) waves from both matter and B-field contributions. 
In general, GWs generated by the magnetic field of this system can propagate with two frequencies $\Omega$ and $2\Omega$ while the matter part only oscillates with one frequency $2\Omega$. 
Also, for $\iota=(n+1)\color{black}\, \pi/2$ ($\iota=n\pi$) in which $n=0,1,2,\dots$, $H_{+\text{F}}$ and $H_{\times\text{F}}$ are the monochromatic waves with the frequencies $2\Omega$ ($2\Omega$) and $\Omega$ ($2\Omega$), respectively. And in the angle $\iota=(n+1)\color{black}\, \pi/2$, the matter part of the cross component completely damps and the corresponding plus polarization slightly oscillates with the frequency $2\Omega$. Furthermore, it has been seen as long as the magnetic axis is parallel to the rotation axis, i.e., $\theta_0=n\pi$ where $n=0,1,2,\dots$, the B-field part of the plus and cross polarizations vanish.

Moreover, we have shown that for all values of $\iota$, the strongest plus polarization occurs at $\theta_0=\pi/2$ and the peaks of the cross polarization move with respect to $\theta_0$ by increasing $\iota$. 
We have specially displayed that there is a condition, i.e., when $\iota=\pi/2$, under which the matter part produces no wave, but the B-field part can emit the GW signals.

In Subsec. \ref{Realistic model}, by considering the realistic model which has a dense rotating magnetosphere, we have concluded that this realistic system can radiate GWs with the same order as the vacuum model. In Subsec. \ref{Application to magnetars}, we have estimated the B-field part amplitude of the millisecond magnetar. The GW amplitudes of the 21 well-known magnetars have been also studied in Table. \ref{table1}. One can see that $h_{0\text{F}}$ of most magnetars is extremely weak and is about $10^{-35}$. 
Finally, by considering the sensitivity of the detectors in the frequency range of $10 \,\text{Hz}<f<10^3 \,\text{Hz}$, we have studied the detectability of the B-field part of GW emitted from the millisecond magnetar in Sec. \ref{Detectors}.

By regarding the results of Sec. \ref{Detectors}, one may conclude that only B-field GW coming from millisecond magnetars with $B=10^{15}\,\text{G}$ located at galactic distances are detectable.
In fact, if the intra-galactic magnetar candidates were born with a very fast rotating period $\sim1\text{ms}$ and dipole B-fields $\sim 10^{15}\,\text{G}$, their magnetic field would emit a strong GW signal so that it could be detected by the third-generation detectors in the first few days after their formation.
However, by considering the millisecond magnetar birth rate within this distance, these systems are less likely to be observed. 

In this paper, we have considered several simple assumptions, e.g., the magnetic field configuration is a tilted dipole, under which the hypothesis of B-field GW emission from a nearby millisecond magnetar can be established. Although we should keep it in mind that more realistic conditions as considering the multipole B-field structure and adding processes that can change the angle $\theta_0$ and subsequently affect the efficiency of the gravitational radiation (\citealp{dall2009early,lander2018neutron}) can provide more accurate results. 

Another point is what could be learned from this type of detection. First, constructing a suitable model consistent with the properties of the signal, like the method introduced in \cite{abbott2017gw170817}, could be used for a single rotating magnetar to measure $\iota$.
Furthermore, utilizing electromagnetic data of millisecond magnetars and measuring $H_{\times}$ and $H_{+}$ of the corresponding gravitational-wave signals, it can be expected that $\iota$ is precisely measurable.
Second, for the case $\iota\simeq\pi/2$, an interesting situation may happen in reality. Assume a nascent magnetar that its intrinsic parameters are known. For typical NS characteristics and $\epsilon_{\text{M}}\simeq 10^{-6}$, it is theoretically expected that this deformed system would be a gravitational wave source with an amplitude of order $\sim 10^{-24}$. On the other hand, in this thought experiment, the detectors just observe a much weaker signal emitted from this system. One may infer from comparing the waveform $H_{\text{F}}$ with $H_\text{M}$, especially for the cross polarizations, that the inclination angle of this is about $\sim \pi/2$, and the detected signal is actually originated from the pure part of the magnetic field. It should be noted that considering $\iota\simeq \pi/2$ is not far from reality and a system with such an inclination angle may be seen in nature (\citealp{ng2008fitting}). However, as we have discussed the event rate of millisecond magnetar at galactic distances is small. Accordingly, the probability of observing $\iota\simeq \pi/2$ would also be small.

Third, by considering the minimum value of the sensitivity of the GW detectors $h_{\text{min}}$ and applying the constrain $h_{\text{c}}<h_{\text{min}}$, one can derive a limit on the millisecond magnetar parameters, especially, on $B$ and $R_{\text{d}}$, for which the detectors cannot monitor the B-field gravitational waves.
In fact, searches for GWs generated by magnetic fields of millisecond magnetar at birth can rule out the possibility of a millisecond magnetar with a super-strong magnetic field.
Finally, considering the simple two-dimensional magnetic fields that are proportional to $\cos n\varphi$, one can estimate that the B-field GW waveform is proportional to $\cos n\left(\Omega t +\varphi \right)$ after applying the definitions \eqref{IjkF} and \eqref{QFjkab}. Here, $n=1,2,3,\cdots$ stands for the dipole, quadrupole, and sextupole magnetic field, respectively. So, knowing the frequency of the rotating magnetars $\Omega$ and comparing it with those of GW signals received, one can deduce the n-value and consequently, a multi-polar field structure of the millisecond magnetars.

As the direct contribution of the magnetic field seems to be impressive, in the sequel, it would be interesting to study the black hole-magnetar and magnetar-magnetar binary systems and evaluate the direct contribution of the magnetic field associated with these systems to the gravitational-wave signal (\citealp{Nazari2020Gravitational}).

\section*{acknowledgments}

We would like to thank Bahram Mashhoon and Nicolas Chamel for reading the manuscript and providing us with constructive comments and helpful references. Worthwhile comments by the anonymous referee are also gratefully acknowledged.

\section*{DATA AVAILABILITY}
No new data were generated in support of this research.
\bibliographystyle{mnras}
\bibliography{short,PN_GWs} 

\newpage
\onecolumn 
\appendix
\section{Supplementary relations and Schemes}\label{AppI}

In order to avoid displaying too many equations, we have extracted the components of the 4-pole-moment tensor $\mathcal{Q}^{jkab}_{\text{F}}$ from the main text of this paper and introduced them in the current appendix.
To obtain the components of $\mathcal{Q}^{jkab}_{\text{F}}$ for the single magnetar, we utilize the same method as mentioned in Subsec. \ref{Simple model}. Hereafter, for the sake of brevity, we will omit the index "$_\text{F}$" and exhibit the 4-pole-moment tensor as $\mathcal{Q}^{jkab}$.  After simplifying and some manipulations, we arrive at
\begin{align}\label{Q^xxxx}
\begin{split}
&\mathcal{Q}^{xxxx}=\frac{A_0}{21}\Big(97+11\cos 2\theta_0-22\sin^2 \theta_0\cos 2\varphi_0 \Big),\\
&\mathcal{Q}^{xxxy}=\frac{8A_0}{7}\sin^2\theta_0\sin 2\varphi_0,\\
&\mathcal{Q}^{xxxz}=\frac{8A_0}{7}\sin 2\theta_0\cos \varphi_0,\\
&\mathcal{Q}^{xxyy}=\frac{A_0}{21}\Big(37-\cos 2\theta_0-70\sin^2\theta_0\cos 2\varphi_0\Big),\\
&\mathcal{Q}^{xxyz}=\frac{12A_0}{7}\sin 2\theta_0\sin\varphi_0,\\
&\mathcal{Q}^{xxzz}=\frac{A_0}{21}\Big(55+53\cos 2\theta_0-34\sin^2\theta_0\cos 2\varphi_0\Big),
\end{split}
\begin{split}
&\mathcal{Q}^{yyxx}=\frac{A_0}{21}\Big(37-\cos 2\theta_0+70\sin^2\theta_0\cos 2\varphi_0 \Big),\\
&\mathcal{Q}^{yyxy}=\frac{8A_0}{7}\sin^2\theta_0\sin 2\varphi_0,\\
&\mathcal{Q}^{yyxz}=\frac{12 A_0}{7}\sin 2\theta_0\cos \varphi_0,\\
&\mathcal{Q}^{yyyy}=\frac{A_0}{21}\Big(97+11\cos 2\theta_0+22\sin^2\theta_0\cos 2\varphi_0\Big),\\
&\mathcal{Q}^{yyyz}=\frac{8A_0}{7}\sin 2\theta_0 \sin\varphi_0,\\
&\mathcal{Q}^{yyzz}=\frac{A_0}{21}\Big(55+53\cos 2\theta_0+34\sin^2\theta_0\cos 2\varphi_0\Big),
\end{split}
\end{align}
for the $x-x-j-k$ and $y-y-j-k$ components. For the $x-y-j-k$ and $x-z-j-k$ components we have
\begin{align}\label{Q^xyxx}
\begin{split}
&\mathcal{Q}^{xyxx}=-\frac{46 A_0}{21}\sin^2 \theta_0\sin 2\varphi_0,\\
&\mathcal{Q}^{xyxy}=\frac{2A_0}{7}\Big(5+\cos 2\theta_0\Big),\\
&\mathcal{Q}^{xyxz}=-\frac{2A_0}{7}\sin 2\theta_0\sin\varphi_0,\\
&\mathcal{Q}^{xyyy}=-\frac{46A_0}{21}\sin^2\theta_0\sin 2\varphi_0,\\
&\mathcal{Q}^{xyyz}=-\frac{2A_0}{7}\sin 2\theta_0\cos\varphi_0,\\
&\mathcal{Q}^{xyzz}=-\frac{34A_0}{21}\sin^2\theta_0\sin 2\varphi_0,
\end{split}
\begin{split}
&\mathcal{Q}^{xzxx}=-\frac{46 A_0}{21}\sin 2\theta_0\cos \varphi_0,\\
&\mathcal{Q}^{xzxy}=-\frac{2A_0}{7}\sin 2\theta_0\sin\varphi_0,\\
&\mathcal{Q}^{xzxz}=\frac{4A_0}{7}\Big(2\cos^2\theta_0+\sin^2\theta_0\big(3-\cos^2\varphi_0\big)\Big),\\
&\mathcal{Q}^{xzyy}=-\frac{34A_0}{21}\sin 2\theta_0\cos\varphi_0,\\
&\mathcal{Q}^{xzyz}=-\frac{4A_0}{7}\sin^2\theta_0\sin\varphi_0\cos\varphi_0,\\
&\mathcal{Q}^{xzzz}=-\frac{46A_0}{21}\sin 2\theta_0\cos \varphi_0.
\end{split}
\end{align}
Similarly for the $y-z-j-k$ and $z-z-j-k$ components in the rotating frame, one may find
\begin{align}\label{Q^yzxx}
\begin{split}
&\mathcal{Q}^{yzxx}=-\frac{34A_0}{21}\sin 2\theta_0\sin\varphi_0,\\
&\mathcal{Q}^{yzxy}=-\frac{2A_0}{7}\sin 2\theta_0\cos\varphi_0,\\
&\mathcal{Q}^{yzxz}=-\frac{4A_0}{7}\sin^2\theta_0\sin\varphi_0\cos\varphi_0,\\
&\mathcal{Q}^{yzyy}=-\frac{46A_0}{21}\sin 2\theta_0\sin\varphi_0,\\
&\mathcal{Q}^{yzyz}=\frac{A_0}{7}\Big(9-\cos 2\theta_0+2\sin^2\theta_0\cos 2\varphi_0\Big),\\
&\mathcal{Q}^{yzzz}=-\frac{46 A_0}{21}\sin 2\theta_0\sin\varphi_0,
\end{split}
\begin{split}
&\mathcal{Q}^{zzxx}=\frac{4A_0}{21}\Big(5-13\cos 2\theta_0+9\sin^2\theta_0\cos 2\varphi_0\Big),\\
&\mathcal{Q}^{zzxy}=\frac{12A_0}{7}\sin^2\theta_0\sin 2\varphi_0,\\
&\mathcal{Q}^{zzxz}=\frac{8A_0}{7}\sin 2\theta_0\cos\varphi_0,\\
&\mathcal{Q}^{zzyy}=\frac{4A_0}{21}\Big(5-13\cos 2\theta_0-9\sin^2\theta_0\cos 2\varphi_0\Big),\\
&\mathcal{Q}^{zzyz}=\frac{8A_0}{7}\sin 2\theta_0\sin\varphi_0,\\
&\mathcal{Q}^{zzzz}=\frac{2 A_0}{21}\Big(43-11\cos 2\theta_0\Big).
\end{split}
\end{align}
It should be mentioned that, as it is obvious from the definition of $\mathcal{Q}^{jkab}_{\text{F}}$, see Eq. \eqref{QFjkab}, this tensor is symmetric under the exchange of the first pair of indices, $j$ and $k$, as well as the second pair of indices, $a$ and $b$. By applying the tensor transformation law \eqref{moment-tensor1}, we obtain the components of the 4-pole-moment tensor in the non-rotating frame as
\begin{align}
\nonumber
&\mathcal{Q}^{x'x'x'x'}=\frac{A_0}{21}\Big(97+11\cos 2\theta_0-22\sin^2 \theta_0\cos 2\left(\Omega t+\varphi_0\right) \Big),\\\nonumber
&\mathcal{Q}^{x'x'x'y'}=\frac{8A_0}{7}\sin^2\theta_0\sin 2\left(\Omega t+\varphi_0\right),\\\nonumber
&\mathcal{Q}^{x'x'x'z'}=\frac{8A_0}{7}\sin 2\theta_0\cos \left(\Omega t+\varphi_0\right),\\
\label{Qx'x'x'x'}
&\mathcal{Q}^{x'x'y'y'}=\frac{A_0}{21}\Big(37-\cos 2\theta_0-70\sin^2\theta_0\cos 2\left(\Omega t+\varphi_0\right)\Big),\\\nonumber
&\mathcal{Q}^{x'x'y'z'}=\frac{12A_0}{7}\sin 2\theta_0\sin\left(\Omega t+\varphi_0\right),\\\nonumber
&\mathcal{Q}^{x'x'z'z'}=\frac{A_0}{21}\Big(55+53\cos 2\theta_0-34\sin^2\theta_0\cos 2\left(\Omega t+\varphi_0\right)\Big),
\end{align}
for the $x'-x'-j'-k'$ components. For $y'-y'-j'-k'$ components, we obtain
\begin{align}
\nonumber
&\mathcal{Q}^{y'y'x'x'}=\frac{A_0}{21}\Big(37-\cos 2\theta_0+70\sin^2\theta_0\cos 2\left(\Omega t+\varphi_0\right) \Big),\\\nonumber
&\mathcal{Q}^{y'y'x'y'}=\frac{8A_0}{7}\sin^2\theta_0\sin 2\left(\Omega t+\varphi_0\right),\\
\label{Qy'y'x'x'}
&\mathcal{Q}^{y'y'x'z'}=\frac{12 A_0}{7}\sin 2\theta_0\cos \left(\Omega t+\varphi_0\right),\\\nonumber
&\mathcal{Q}^{y'y'y'y'}=\frac{A_0}{21}\Big(97+11\cos 2\theta_0+22\sin^2\theta_0\cos 2\left(\Omega t+\varphi_0\right)\Big),\\\nonumber
&\mathcal{Q}^{y'y'y'z'}=\frac{8A_0}{7}\sin 2\theta_0 \sin\left(\Omega t+\varphi_0\right),\\\nonumber
&\mathcal{Q}^{y'y'z'z'}=\frac{A_0}{21}\Big(55+53\cos 2\theta_0+34\sin^2\theta_0\cos 2\left(\Omega t+\varphi_0\right)\Big).
\end{align}
Moreover, the $x'-y'-j'-k'$ and $x'-z'-j'-k'$ components are
\begin{align}\label{Qx'y'x'x'}
\begin{split}
&\mathcal{Q}^{x'y'x'x'}=-\frac{46 A_0}{21}\sin^2 \theta_0\sin 2\left(\Omega t+\varphi_0\right),\\
&\mathcal{Q}^{x'y'x'y'}=\frac{2A_0}{7}\Big(5+\cos 2\theta_0\Big),\\
&\mathcal{Q}^{x'y'x'z'}=-\frac{2A_0}{7}\sin 2\theta_0\sin\left(\Omega t+\varphi_0\right),\\
&\mathcal{Q}^{x'y'y'y'}=-\frac{46A_0}{21}\sin^2\theta_0\sin 2\left(\Omega t+\varphi_0\right),\\
&\mathcal{Q}^{x'y'y'z'}=-\frac{2A_0}{7}\sin 2\theta_0\cos\left(\Omega t+\varphi_0\right),\\
&\mathcal{Q}^{x'y'z'z'}=-\frac{34A_0}{21}\sin^2\theta_0\sin 2\left(\Omega t+\varphi_0\right),
\end{split}
\begin{split}
&\mathcal{Q}^{x'z'x'x'}=-\frac{46 A_0}{21}\sin 2\theta_0\cos \left(\Omega t+\varphi_0\right),\\
&\mathcal{Q}^{x'z'x'y'}=-\frac{2A_0}{7}\sin 2\theta_0\sin\left(\Omega t+\varphi_0\right),\\
&\mathcal{Q}^{x'z'x'z'}=\frac{4A_0}{7}\Big(2\cos^2\theta_0+\sin^2\theta_0\big(3-\cos^2\left(\Omega t+\varphi_0\right)\big)\Big),\\
&\mathcal{Q}^{x'z'y'y'}=-\frac{34A_0}{21}\sin 2\theta_0\cos\left(\Omega t+\varphi_0\right),\\
&\mathcal{Q}^{x'z'y'z'}=-\frac{4A_0}{7}\sin^2\theta_0\sin\left(\Omega t+\varphi_0\right)\cos\left(\Omega t+\varphi_0\right),\\
&\mathcal{Q}^{x'z'z'z'}=-\frac{46A_0}{21}\sin 2\theta_0\cos \left(\Omega t+\varphi_0\right).
\end{split}
\end{align}
And finally, we find the $y'-z'-j'-k'$ and $z'-z'-j'-k'$ components as follows
\begin{align}\label{Qz'z'x'x'}
\begin{split}
&\mathcal{Q}^{y'z'x'x'}=-\frac{34A_0}{21}\sin 2\theta_0\sin\left(\Omega t+\varphi_0\right),\\
&\mathcal{Q}^{y'z'x'y'}=-\frac{2A_0}{7}\sin 2\theta_0\cos\left(\Omega t+\varphi_0\right),\\
&\mathcal{Q}^{y'z'x'z'}=-\frac{4A_0}{7}\sin^2\theta_0\sin\left(\Omega t+\varphi_0\right)\cos\left(\Omega t+\varphi_0\right),\\
&\mathcal{Q}^{y'z'y'y'}=-\frac{46A_0}{21}\sin 2\theta_0\sin\left(\Omega t+\varphi_0\right),\\
&\mathcal{Q}^{y'z'y'z'}=\frac{A_0}{7}\Big(9-\cos 2\theta_0+2\sin^2\theta_0\cos 2\left(\Omega t+\varphi_0\right)\Big),\\
&\mathcal{Q}^{y'z'z'z'}=-\frac{46 A_0}{21}\sin 2\theta_0\sin\left(\Omega t+\varphi_0\right),
\end{split}
\begin{split}
&\mathcal{Q}^{z'z'x'x'}=\frac{4A_0}{21}\Big(5-13\cos 2\theta_0+9\sin^2\theta_0\cos 2\left(\Omega t+\varphi_0\right)\Big),\\
&\mathcal{Q}^{z'z'x'y'}=\frac{12A_0}{7}\sin^2\theta_0\sin 2\left(\Omega t+\varphi_0\right),\\
&\mathcal{Q}^{z'z'x'z'}=\frac{8A_0}{7}\sin 2\theta_0\cos\left(\Omega t+\varphi_0\right),\\
&\mathcal{Q}^{z'z'y'y'}=\frac{4A_0}{21}\Big(5-13\cos 2\theta_0-9\sin^2\theta_0\cos 2\left(\Omega t+\varphi_0\right)\Big),\\
&\mathcal{Q}^{z'z'y'z'}=\frac{8A_0}{7}\sin 2\theta_0\sin\left(\Omega t+\varphi_0\right),\\
&\mathcal{Q}^{z'z'z'z'}=\frac{2 A_0}{21}\Big(43-11\cos 2\theta_0\Big).
\end{split}
\end{align}
This is obvious in the sense that this coordinate transformation, like transformation of $\mathcal{I}^{jk}$, is equivalent to the following change: $\varphi_0\rightarrow (\Omega t+\varphi_0)$ in the $\mathcal{Q}^{jkab}$ components. As the next step, by evaluating the second time derivatives of the 4-pole-moment tensor, we obtain its components as
\begin{align}\label{Qxxzz}
\begin{split}
&\ddot{\mathcal{Q}}^{x'x'x'x'}= \frac{88A_0\Omega^2}{21}\sin^2\theta_0\cos 2\left(\Omega t+\varphi_0\right),\\
&\ddot{\mathcal{Q}}^{x'x'x'y'}=-\frac{32A_0\Omega^2}{7}\sin^2\theta_0\sin 2\left(\Omega t+\varphi_0\right),\\
&\ddot{\mathcal{Q}}^{x'x'x'z'}=-\frac{8A_0\Omega^2}{7}\sin 2\theta_0\cos\left(\Omega t+\varphi_0\right),\\
&\ddot{\mathcal{Q}}^{x'x'y'y'}=\frac{40A_0\Omega^2}{3}\sin^2\theta_0\cos 2\left(\Omega t+\varphi_0\right),\\
&\ddot{\mathcal{Q}}^{x'x'y'z'}=-\frac{12A_0\Omega^2}{7}\sin 2\theta_0\sin \left(\Omega t+\varphi_0\right),\\
&\ddot{\mathcal{Q}}^{x'x'z'z'}=\frac{136A_0\Omega^2}{21}\sin^2\theta_0\cos 2\left(\Omega t+\varphi_0\right),
\end{split}
\begin{split}
&\ddot{\mathcal{Q}}^{y'y'x'x'}=-\frac{40A_0\Omega^2}{3}\sin^2\theta_0\cos 2\left(\Omega t+\varphi_0\right),\\
&\ddot{\mathcal{Q}}^{y'y'x'y'}=-\frac{32A_0\Omega^2}{7}\sin^2\theta_0\sin 2\left(\Omega t+\varphi_0\right),\\
&\ddot{\mathcal{Q}}^{y'y'x'z'}=-\frac{12A_0\Omega^2}{7}\sin 2\theta_0\cos\left(\Omega t+\varphi_0\right),\\
&\ddot{\mathcal{Q}}^{y'y'y'y'}=-\frac{88A_0\Omega^2}{21}\sin^2\theta_0\cos 2\left(\Omega t+\varphi_0\right),\\
&\ddot{\mathcal{Q}}^{y'y'y'z'}=-\frac{8A_0\Omega^2}{7}\sin 2\theta_0\sin \left(\Omega t+\varphi_0\right),\\
&\ddot{\mathcal{Q}}^{y'y'z'z'}=-\frac{136A_0\Omega^2}{21}\sin^2\theta_0\cos 2\left(\Omega t+\varphi_0\right),
\end{split}
\end{align}
for $\ddot{\mathcal{Q}}^{x'x'j'k'}$ and $\ddot{\mathcal{Q}}^{y'y'j'k'}$,
\begin{align}\label{Qxyzz}
\begin{split}
&\ddot{\mathcal{Q}}^{x'y'x'x'}=\frac{184A_0\Omega^2}{21}\sin^2\theta_0\sin 2\left(\Omega t+\varphi_0\right),\\
&\ddot{\mathcal{Q}}^{x'y'x'y'}= 0,\\
&\ddot{\mathcal{Q}}^{x'y'x'z'}=\frac{2A_0\Omega^2}{7}\sin 2\theta_0\sin\left(\Omega t+\varphi_0\right),\\
&\ddot{\mathcal{Q}}^{x'y'y'y'}=\frac{184A_0\Omega^2}{21}\sin^2\theta_0\sin 2\left(\Omega t+\varphi_0\right),\\
&\ddot{\mathcal{Q}}^{x'y'y'z'}=\frac{2A_0\Omega^2}{7}\sin 2\theta_0\cos \left(\Omega t+\varphi_0\right),\\
&\ddot{\mathcal{Q}}^{x'y'z'z'}=\frac{136A_0\Omega^2}{21}\sin^2\theta_0\sin 2\left(\Omega t+\varphi_0\right),
\end{split}
\begin{split}
&\ddot{\mathcal{Q}}^{x'z'x'x'}=\frac{46A_0\Omega^2}{21}\sin 2\theta_0\cos \left(\Omega t+\varphi_0\right),\\
&\ddot{\mathcal{Q}}^{x'z'x'y'}=\frac{2A_0\Omega^2}{7}\sin 2\theta_0\sin\left(\Omega t+\varphi_0\right),\\
&\ddot{\mathcal{Q}}^{x'z'x'z'}=\frac{8A_0\Omega^2}{7}\sin^2\theta_0\cos 2\left(\Omega t+\varphi_0\right),\\
&\ddot{\mathcal{Q}}^{x'z'y'y'}=\frac{34A_0\Omega^2}{21}\sin 2\theta_0\cos \left(\Omega t+\varphi_0\right),\\
&\ddot{\mathcal{Q}}^{x'z'y'z'}=\frac{8A_0\Omega^2}{7}\sin^2\theta_0\sin 2\left(\Omega t+\varphi_0\right),\\
&\ddot{\mathcal{Q}}^{x'z'z'z'}=\frac{46A_0\Omega^2}{21}\sin 2\theta_0\cos \left(\Omega t+\varphi_0\right),
\end{split}
\end{align}
for $\ddot{\mathcal{Q}}^{x'y'j'k'}$ and $\ddot{\mathcal{Q}}^{x'z'j'k'}$, and
\begin{align}\label{Qyzzz}
\begin{split}
&\ddot{\mathcal{Q}}^{y'z'x'x'}=\frac{34A_0\Omega^2}{21}\sin 2\theta_0\sin\left(\Omega t+\varphi_0\right),\\
&\ddot{\mathcal{Q}}^{y'z'x'y'}=\frac{2A_0\Omega^2}{7}\sin 2\theta_0\cos\left(\Omega t+\varphi_0\right),\\
&\ddot{\mathcal{Q}}^{y'z'x'z'}=\frac{8A_0\Omega^2}{7}\sin^2\theta_0\sin 2\left(\Omega t+\varphi_0\right),\\
&\ddot{\mathcal{Q}}^{y'z'y'y'}=\frac{46A_0\Omega^2}{21}\sin 2\theta_0\sin \left(\Omega t+\varphi_0\right),\\
&\ddot{\mathcal{Q}}^{y'z'y'z'}=-\frac{8A_0\Omega^2}{7}\sin^2\theta_0\cos 2\left(\Omega t+\varphi_0\right),\\
&\ddot{\mathcal{Q}}^{y'z'z'z'}=\frac{46A_0\Omega^2}{21}\sin 2\theta_0\sin \left(\Omega t+\varphi_0\right),
\end{split}
\begin{split}
&\ddot{\mathcal{Q}}^{z'z'x'x'}=-\frac{48A_0\Omega^2}{7}\sin^2\theta_0\cos 2(\Omega t+\varphi_0),\\
&\ddot{\mathcal{Q}}^{z'z'x'y'}=-\frac{48 A_0 \Omega^2}{7}\sin^2\theta_0\sin 2(\Omega t+\varphi_0),\\
&\ddot{\mathcal{Q}}^{z'z'x'z'}=-\frac{8 A_0\Omega^2}{7}\sin 2\theta_0\cos(\Omega t+\varphi_0),\\
&\ddot{\mathcal{Q}}^{z'z'y'y'}=\frac{48 A_0 \Omega^2}{7}\sin^2\theta_0\cos 2(\Omega t+\varphi_0),\\
&\ddot{\mathcal{Q}}^{z'z'y'z'}=-\frac{8 A_0 \Omega^2}{7}\sin 2\theta_0\sin (\Omega t+\varphi_0),\\
&\ddot{\mathcal{Q}}^{z'z'z'z'}=0,
\end{split}
\end{align}
for $\ddot{\mathcal{Q}}^{y'z'j'k'}$ and $\ddot{\mathcal{Q}}^{z'z'j'k'}$.

\label{lastpage}

\end{document}